\documentclass{aa}  

\usepackage{graphicx}
\usepackage{txfonts}
\usepackage{lscape}
\begin{document} 

   \title{Zeeman-Doppler imaging of five young solar-type stars \thanks{Based on observations made with the HARPSpol instrument on the ESO 3.6 m telescope in La Silla (Chile), under programme IDs 091.D-0836 and 0100.D-0176.}}

   \author{T. Willamo
          \inst{1}
          \and
          J. J. Lehtinen\inst{2}$^,$\inst{1}
          \and
          T. Hackman\inst{1}
          \and
          M. J. Käpylä\inst{3}$^,$ \inst{4}$^,$ \inst{5}
          \and
          O. Kochukhov\inst{6}
          \and
          S. V. Jeffers\inst{4}
          \and
          H. Korhonen\inst{7}
          \and
          S. C. Marsden\inst{8}
          }

   \institute{Department of Physics, P.O. Box 64, FI-00014 University of Helsinki, Finland \\
              \email{teemu.willamo@helsinki.fi}
             \and
             Finnish Centre for Astronomy with ESO (FINCA), University of Turku, Vesilinnantie 5, FI-20014 University of Turku, Finland
             \and 
             Department of Computer Science, Aalto University, PO Box 15400, FI-00076 Aalto, Finland
             \and
             Max Planck Institute for Solar System Research, Justus-von-Liebig-Weg 3, D-37077 Göttingen, Germany
             \and
             Nordita, KTH Royal Institute of Technology and Stockholm University, Hannes Alfv\'ens v\"ag 12, SE-10691 Stockholm, Sweden
             \and
             Department of Physics and Astronomy, Uppsala University, Box 516, 751 20 Uppsala, Sweden
             \and
             European Southern Observatory (ESO), Alonso de C\'{o}rdova 3107, Vitacura, Santiago, Chile
             \and
             University of Southern Queensland, Centre for Astrophysics, Toowoomba, QLD, 4350, Australia
             }

   \date{Received June 28, 2021; accepted November 27, 2021}

 
  \abstract
   {The magnetic activity of the Sun changes with the solar cycle. Similar cycles are found in other stars as well, but their details are not known to a similar degree. Characterising stellar magnetic cycles is important for the understanding of the stellar and solar dynamos that are driving the magnetic activity.}
   {We present spectropolarimetric observations of five young, solar-type stars and compare them to previous observations, with the aim to identify and characterise stellar equivalents of the solar cycle.}
   {We use Zeeman-Doppler imaging (ZDI) to map the surface magnetic field and brightness of our targets. The magnetic field is decomposed into spherical harmonic expansions, from which we report the strengths of the axisymmetric versus non-axisymmetric and poloidal versus toroidal components, and we compare them to the Rossby numbers of the stars.}
   {We present five new ZDI maps of young, solar-type stars from December 2017. Of special interest is the case of V1358 Ori, which had gone through a polarity reversal between our observations and earlier ones. A less evident polarity reversal might also have occurred in HD 35296. There is a preference for a more axisymmetric field, and possibly a more toroidal field, for the more active stars with lower Rossby number, but a larger sample should be studied to draw any strong conclusions from this. For most of the individual stars, the amounts of toroidal and poloidal field have stayed on levels similar  to those in earlier observations.}
   {We find evidence for a magnetic polarity reversal having occurred in V1358 Ori. An interesting target for future observations is $\chi^1$ Ori, which may have a short magnetic cycle of a few years. The correlation between the brightness maps and the magnetic field is mostly poor, which could indicate the presence of small-scale magnetic features of different polarities that cancel one another out and are thus not resolved in our maps.}


   \maketitle
%

\section{Introduction}

Magnetic fields play a crucial part in stellar physics. They cause different kinds of activity phenomena that are observed in the Sun and other stars. In young stars, rapid rotation typically makes them much more magnetically active than the Sun. On the other hand, in older stars of ages comparable to the Sun, the stellar winds have carried away a large part of the angular momentum through magnetic braking, making them slower rotators and thus resulting in lower activity levels.

The magnetic activity of the Sun is driven by a dynamo mechanism. Here a poloidal magnetic field is turned into a toroidal one through differential rotation ($\Omega$ effect) and then again back to poloidal through the $\alpha$ effect. The mechanism behind how the $\alpha$ effect operates is still under debate. Some believe it to work through the Coriolis force, twisting convective bubbles \cite[Parker's $\alpha$ effect;][]{parker1955}, while others believe it to work through rising flux tubes and their decay \cite[Babcock-Leighton mechanism;][]{babcock1961,leighton1964,leighton1969}. In rapidly rotating young stars, with higher levels of magnetic activity than the Sun, the $\Omega$ effect is of lesser importance, and the $\alpha$ effect is believed to drive the transition from toroidal to poloidal field as well, in what is referred to as an $\alpha^2\Omega$ or $\alpha^2$ dynamo. Recent numerical simulations are available in, for example, \cite{warnecke2020}. A review of the dynamo theory is available in, for example, \cite{charbonneau2010}.

The solar dynamo drives the 22 year solar magnetic cycle, where the large-scale solar magnetic field goes through a polarity reversal around the activity maximum, on average every 11th year. The time of these polarity reversals can be defined from the radial magnetic field on polar regions. These changes in the solar magnetic field are the drivers of the 11 year sunspot cycle. Similar cyclicity is observed in other stars as well, through photometry \citep[e.g.][]{lehtinen16} and chromospheric emission \citep[e.g.][]{baliunas1995,olspert2018}. However, in order to compare the magnetic properties -- such as the field strengths, polarities, or topologies of these stellar cycles -- to the solar cycle, more sophisticated methods are needed to resolve and map the stellar magnetic field. Evolution of the magnetic field properties, over the activity cycles, could tell us about the similarities and differences between the solar and stellar dynamos.

Zeeman-Doppler imaging (ZDI) is a powerful tool for mapping the distribution of the large-scale magnetic field on stellar surfaces \citep{semel89,brown1991}. This method allows us to determine the structure and orientation of the large-scale magnetic field. Of special interest are the cases where the polarity of the magnetic field is reversed between ZDI maps from different epochs, possibly in a similar manner as the polarity reversals in the solar cycle. With frequent enough mapping over long time periods, the stellar counterparts of the solar cycle can be characterised.
Such polarity reversals have been detected in LQ Hya \citep{lehtinen2019arXiv}, $\chi^1$ Ori \citep{rosen16}, HD 29615 \citep{waite15,hackman16}, $\tau$ Boo \citep{donati2008_tauboo,fares2009_tauboo,fares2013_tauboo,mengel2016_tauboo,jeffers2018_tauboo}, 61 Cyg A \citep{boro_saikia2016,boro_saikia2018}, HD 75332 \citep{brown2021_hd75332}, $\epsilon$ Eri \citep{jeffers2014_epseri,jeffers2017_epseri}, and $\kappa$ Cet \citep{boro_saikia2021arxiv}. These stars are dwarfs ranging from spectral classes K5 (61 Cyg A) to F8 ($\tau$ Boo). With suitable photometric or chromospheric data, the polarity reversals can be related to the corresponding phases of the activity cycle.

Here we apply ZDI to five young solar-type stars: BE Cet, $\chi^1$ Ori, HD 29615, HD 35296, and V1358 Ori, with the aim to compare their magnetic topologies to previous similar studies. In this way we aim to follow their magnetic evolution on timescales comparable to that of the solar cycle.

\section{Observations}

The Stokes I and V spectropolarimetry was collected with the HARPS\-pol high-resolution spectropolarimeter \citep{piskunov2011_harps}, which is mounted at the ESO 3.6 m telescope in La Silla, Chile. The spectral resolution of HARPS is around $R \approx$ 110 000. The observations, spanning seven nights in December 2017, were part of our programme \textquoteleft Active Suns Revisited\textquoteright. In addition, we re-analysed one dataset of V1358 Ori from 2013, originally published by \cite{hackman16}, since they used an incorrect rotation period.

Each exposure comprises four sub-exposures, where the quarter-wave plate is rotated by 90$^{\circ}$ in between each sub-exposure. 
In addition to the Stokes IV spectra, we also derive a \textquoteleft null spectrum' for each observation, which should contain only noise and can thus be used to check that no spurious signal is included in the line profile. 
The data were reduced with the REDUCE package \citep{piskunov02}. 

The observations are presented in more detail in Table \ref{obs} and summarised for each star in Table \ref{obs2}. The rotational phase coverage, $f_{\phi}$, was estimated by assuming that each spectrum at phase $\phi_{\mathrm{rot}}$ covers a phase range from $\phi_{\mathrm{rot}} - 0.05$ to $\phi_{\mathrm{rot}} + 0.05$. For all the new datasets here, the phase coverage is satisfactory (the worst is 0.543 for $\chi^1$ Ori), while for the 2013 data for V1358 Ori it is quite poor, only 0.4.

As the polarisation signal in individual spectral lines, caused by the magnetic field, is too weak to be reliably detected, we used the least-square deconvolution (LSD) method to enhance the signal. The method works by combining thousands of spectral lines into a mean LSD profile \citep{kochukhov2010_lsd}. The number of spectral lines used for each star is shown in Table \ref{obs2}, as is the resulting mean signal-to-noise ratio, $\langle S/N \rangle$, of the LSD profiles.
The line mask and line parameters were extracted from the Vienna Atomic Line Database (VALD)\footnote{\url{http://vald.astro.uu.se/}} \citep{piskunov95,kupka99}. 
The LSD profiles of all stars are shown in Appendix \ref{profs}, and the null profiles in Appendix \ref{append2}.

\begin{figure*}
   \centering
   \includegraphics[width=6cm]{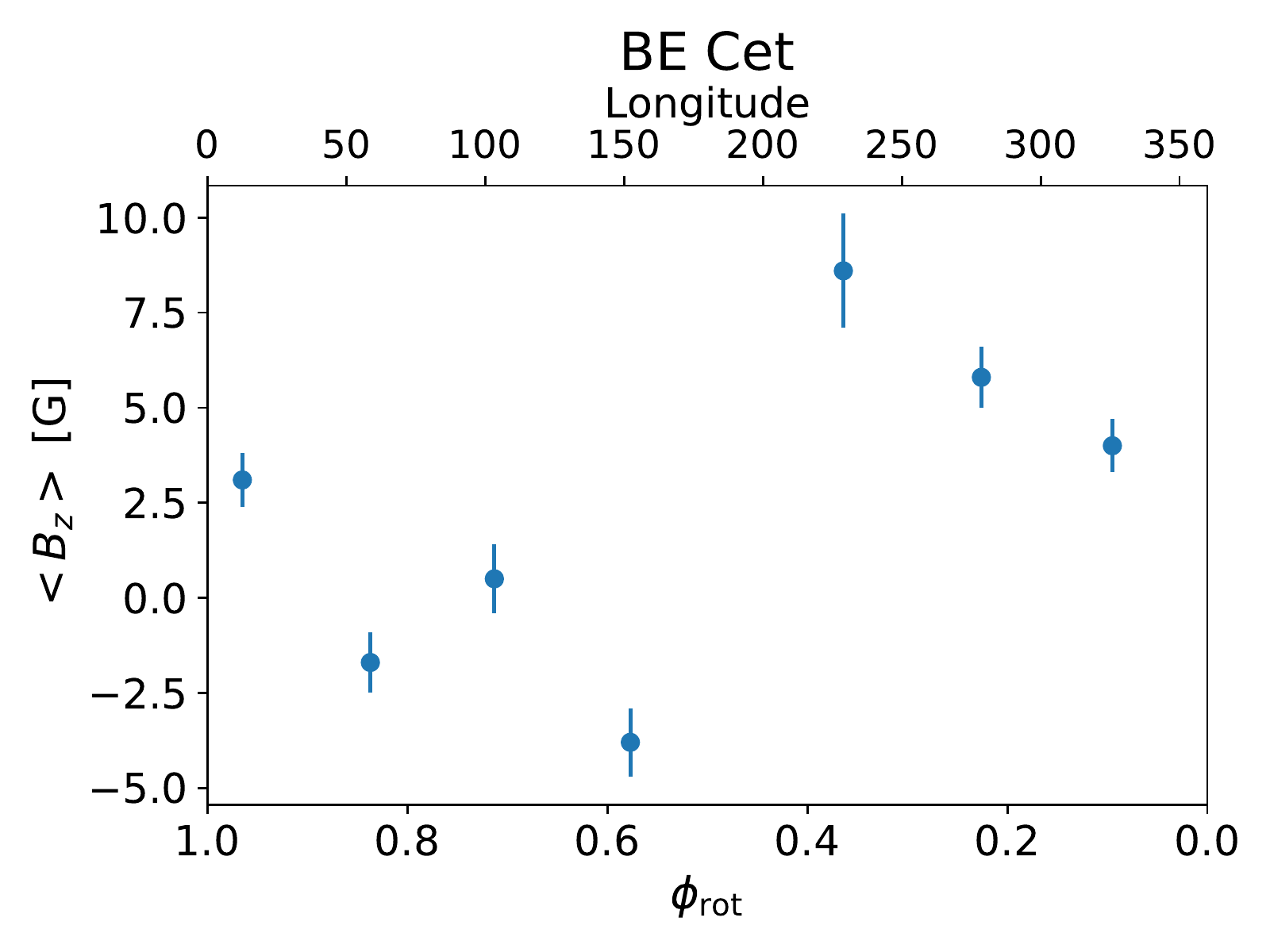}
   \includegraphics[width=6cm]{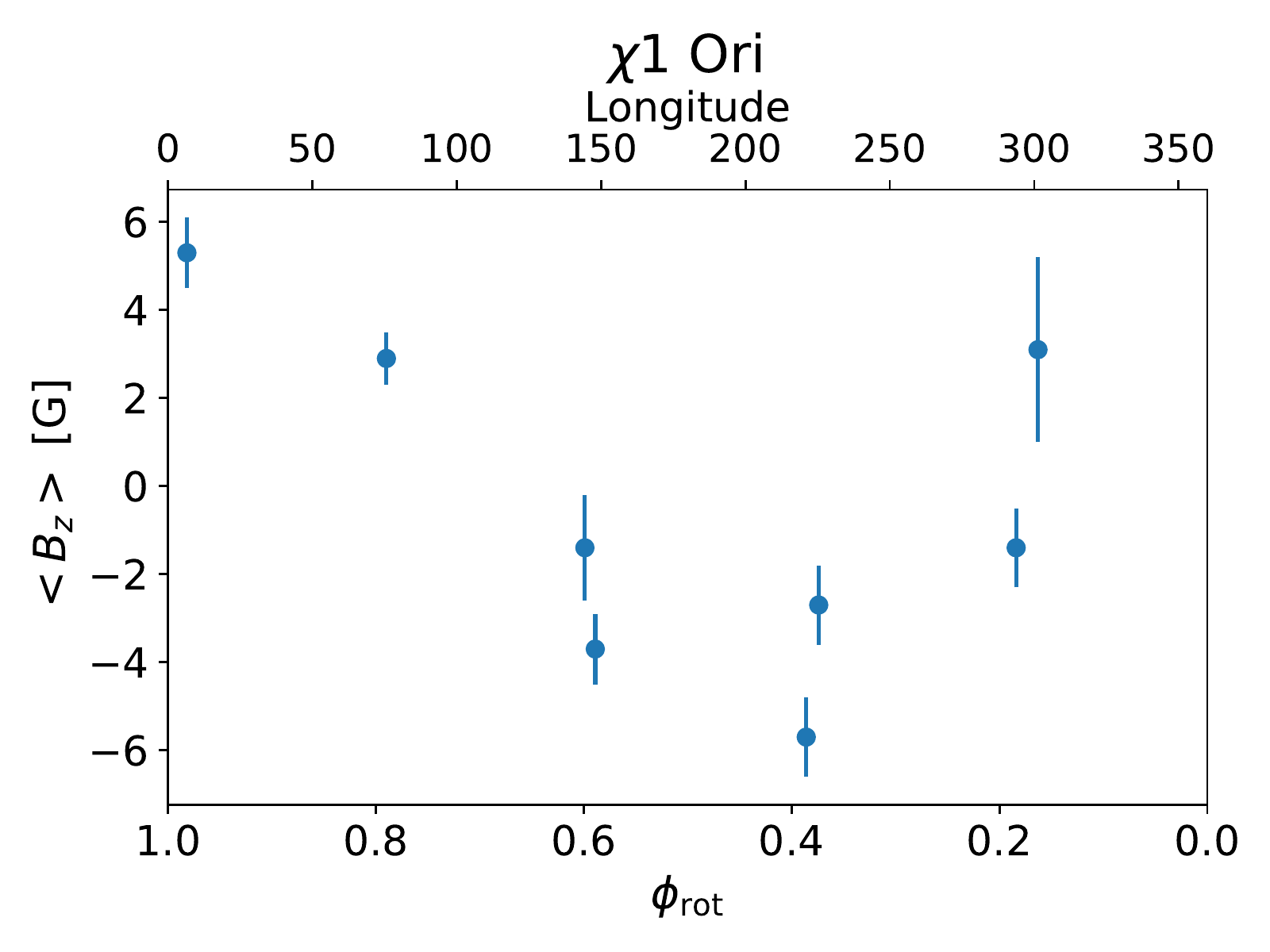}
   \includegraphics[width=6cm]{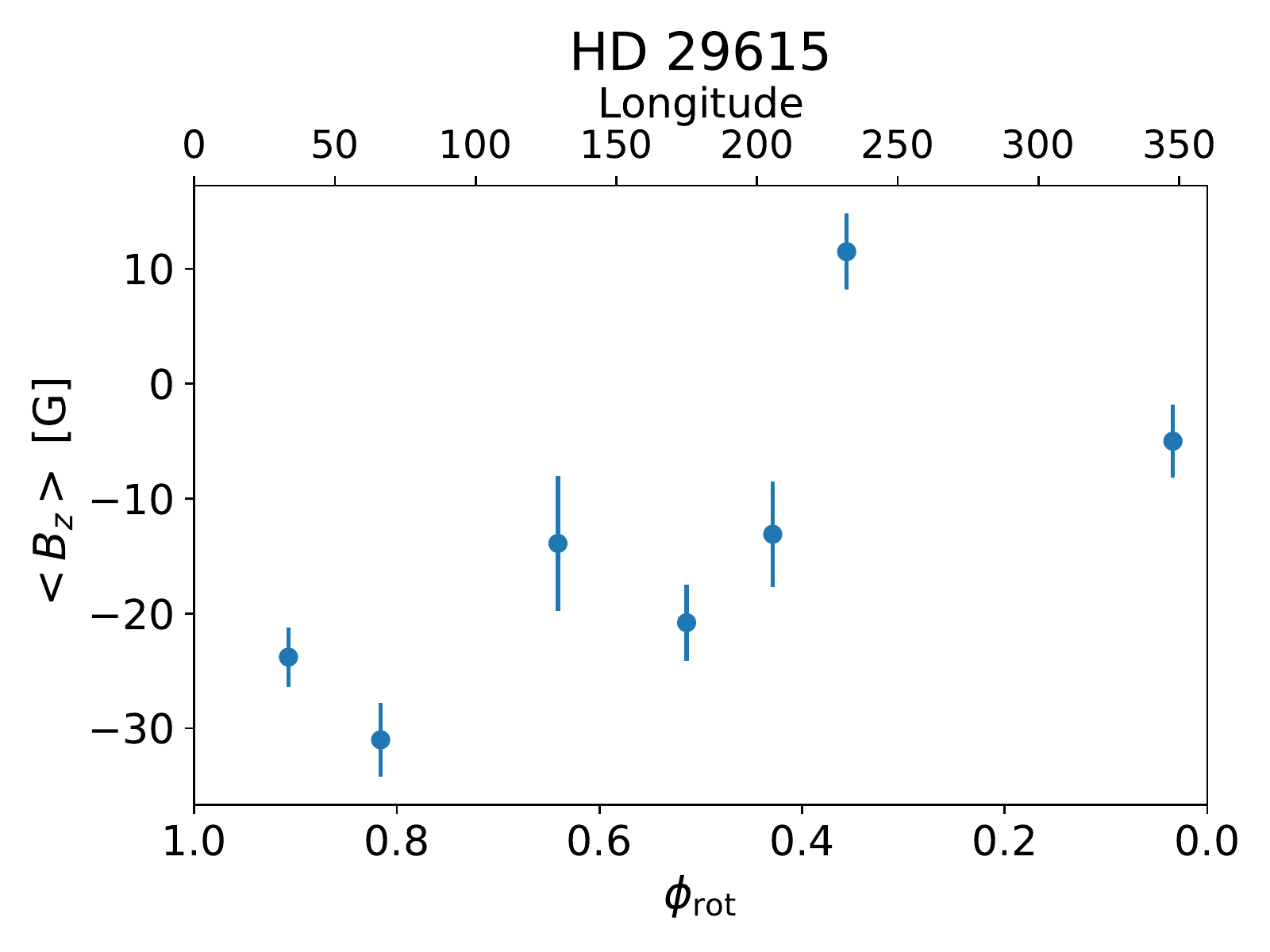}
   \includegraphics[width=6cm]{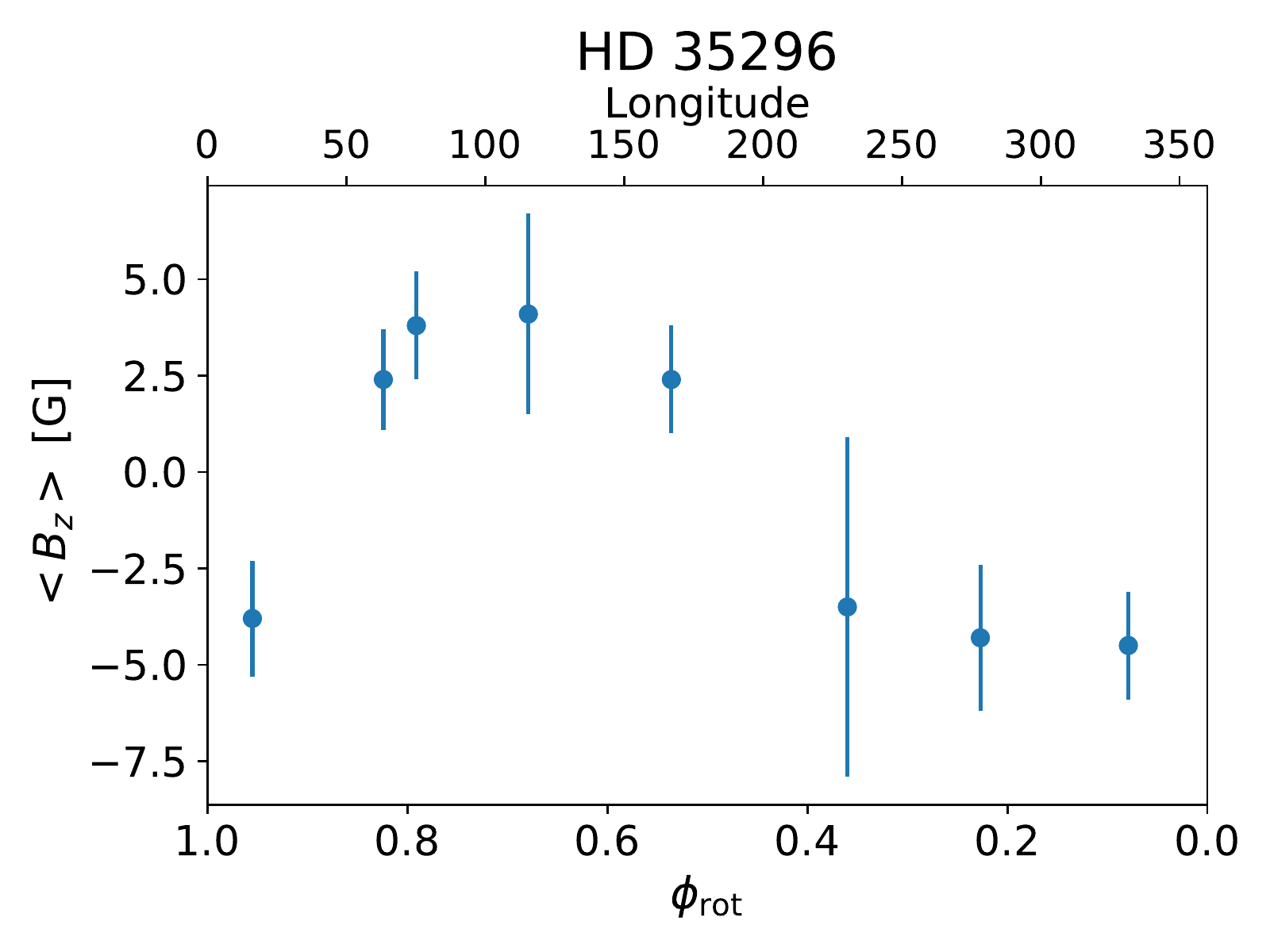}
   \includegraphics[width=6cm]{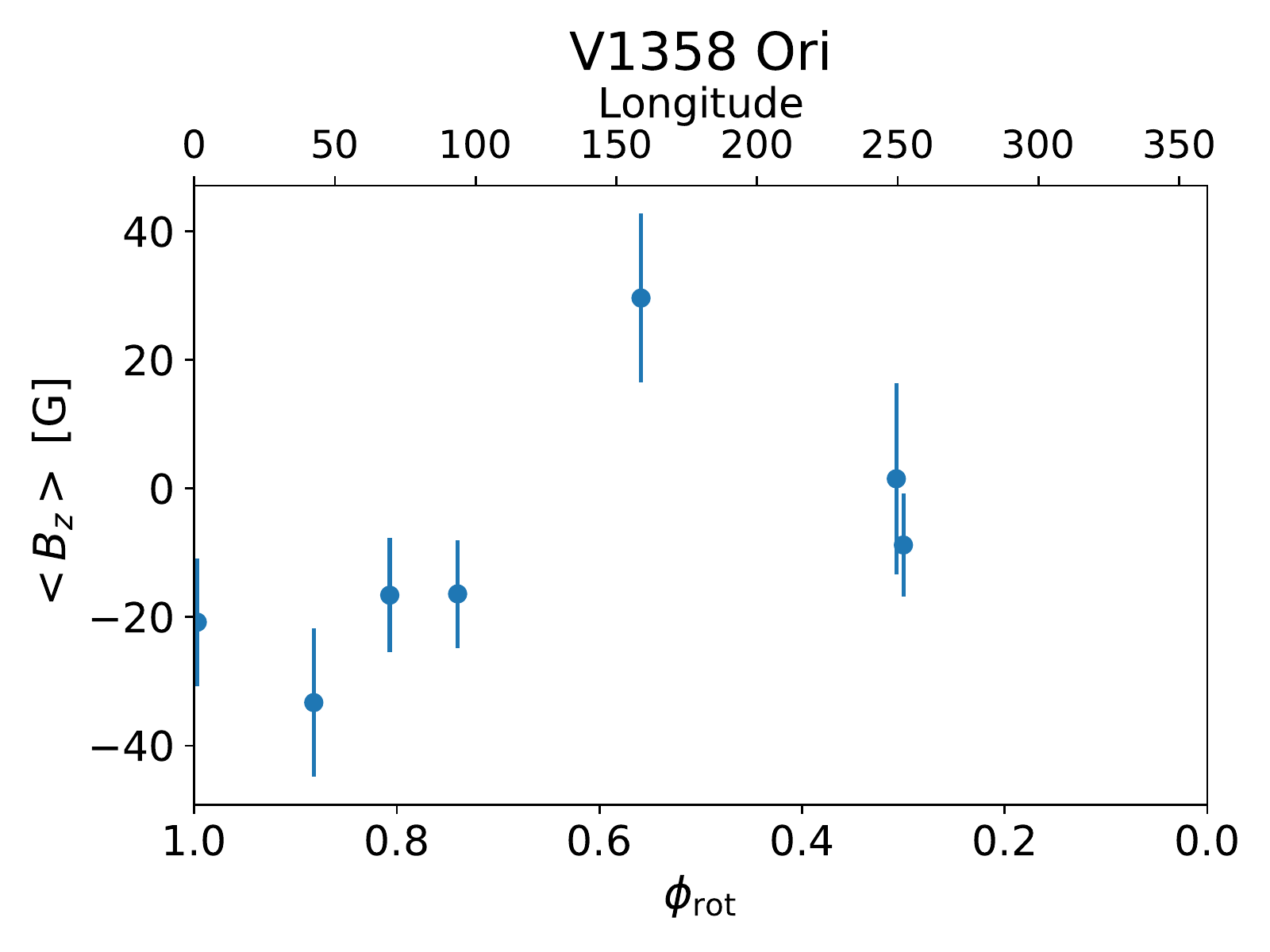}
   \caption{Mean line-of-sight magnetic field, $\langle B_z \rangle$, of the observed stars as a function of their rotational phase. The corresponding longitudes on the stellar surface (which is turned towards the observer at that phase) are shown on the top axis. The phases are inverted in order for the longitudes to match those used in the ZDI maps (Figs. \ref{becet}--\ref{v1358ori}), where the direction of the rotation is chosen to be similar to the Earth's rotation.}
   \label{bz}
\end{figure*}

We also calculated the mean line-of-sight component of the magnetic field, $\langle B_z \rangle$ \citep[see][]{kochukhov2010_lsd}, for all Stokes V profiles and null profiles. They are listed in Table \ref{obs} and shown in Fig. \ref{bz} as a function of the rotational phase for the V profiles. The magnetic field is detected for all our stars, and there are clear variations depending on the phase. However, for V1358 Ori, the faintest of our stars, there are higher signals in the null profiles (higher values in the last column in Table \ref{obs}) than for the other stars. We deem this to be caused by a higher noise level, and it should not significantly affect our results.

We calculated the false alarm probability (FAP) for the magnetic field detection following the method of \cite{donati1992}. Here, the FAP is calculated as

  \begin{equation}
    \mathrm{FAP} = 1-\Gamma\Big(\frac{\nu}{2}, \frac{\nu\chi^2}{2}\Big),
\end{equation}

\noindent where $\Gamma$ is the incomplete gamma function and $\nu$ is the number of resolved elements in the spectrum, which can be calculated as

  \begin{equation}
    \nu = \frac{R(v_2 - v_1)}{c},
  \end{equation}

\noindent where $R$ is the spectral resolution (110 000 for HARPS), $c$ is the speed of light, and $v_2 - v_1$ is the radial velocity interval of the line profile. The $\chi^2$ test statistics are calculated as

  \begin{equation}
    \chi^2 = \frac{1}{n} \sum_{i=1}^{n}\Big(\frac{V(\lambda_{\mathrm{i}})}{\sigma_{\mathrm{i}}}\Big)^2,
  \end{equation}

\noindent where $V$ is the Stokes V profile or null profile, $\sigma_{\mathrm{i}}$ its uncertainty, and $\lambda_{\mathrm{i}}$ the wavelength. These FAP values are included in Table \ref{obs}, both for the V profiles and the null profiles. We note again that the values for V1358 Ori are significantly larger than for the rest of the stars due to a higher noise level. There is, nevertheless, at least one definite magnetic field detection among the profiles for V1358 Ori (requiring a 3 $\sigma$ certainty, or FAP < 0.0027), which should allow for meaningful magnetic mapping.

\begin{table*}
\centering
\caption{\label{obs}Observation log.}
\begin{tabular}{c c c c c c c c}
\hline\hline
Date (UT) & HJD-2458000 & $\phi_{\mathrm{rot}}$ & $S/N$ & $\langle B_z \rangle $ [G] & Null $\langle B_z \rangle$ [G] & FAP [V] & FAP [null] \\
\hline
BE Cet & & & & & & & \\
\hline
2017 12 13 1:30 & 100.563 & 0.577 & 17205 & -3.8 $\pm$ 0.9 & 0.9 $\pm$ 0.9 & $<10^{-16}$ & 0.66 \\
2017 12 14 2:34 & 101.608 & 0.713 & 16530 & 0.5 $\pm$ 0.9 & -0.0 $\pm$ 0.9 & 2.99 $\times 10^{-5}$ & 0.91 \\
2017 12 15 1:17 & 102.554 & 0.837 & 19330 & -1.7 $\pm$ 0.8 & 0.0 $\pm$ 0.8 & $<10^{-16}$ & 0.83 \\
2017 12 16 0:56 & 103.539 & 0.965 & 20295 & 3.1 $\pm$ 0.7 & -0.1 $\pm$ 0.7 & $<10^{-16}$ & 0.84 \\
2017 12 17 0:56 & 104.539 & 0.095 & 22277 & 4.0 $\pm$ 0.7 & -0.2 $\pm$ 0.7 & 1.46 $\times 10^{-13}$ & 0.84 \\
2017 12 18 0:55 & 105.539 & 0.226 & 19332 & 5.8 $\pm$ 0.8 & -1.3 $\pm$ 0.8 & 2.57 $\times 10^{-10}$ & 0.91 \\
2017 12 19 2:31 & 106.605 & 0.364 & 9861 & 8.6 $\pm$ 1.5 & -1.2 $\pm$ 1.5 & $<10^{-16}$ & 0.74 \\
\hline
$\chi$1 Ori & & & & & & & \\
\hline
2017 12 13 5:20 & 100.728 & 0.184 & 17534 & -1.4 $\pm$ 0.9 & -0.0 $\pm$ 0.9 & 4.45 $\times 10^{-8}$ & 0.88 \\
2017 12 14 4:12 & 101.681 & 0.374 & 17732 & -2.7 $\pm$ 0.9 & 0.6 $\pm$ 0.9 & 9.05 $\times 10^{-3}$ & 0.96 \\
2017 12 15 6:06 & 102.760 & 0.589 & 20217 & -3.7 $\pm$ 0.8 & -0.9 $\pm$ 0.8 & 1.73 $\times 10^{-10}$ & 0.76 \\
2017 12 15 7:19 & 102.811 & 0.599 & 13244 & -1.4 $\pm$ 1.2 & 0.1 $\pm$ 1.2 & 1.74 $\times 10^{-3}$ & 0.74 \\
2017 12 16 6:19 & 103.769 & 0.790 & 24453 & 2.9 $\pm$ 0.6 & -0.2 $\pm$ 0.6 & 7.02 $\times 10^{-6}$ & 0.83 \\
2017 12 17 5:30 & 104.735 & 0.982 & 20513 & 5.3 $\pm$ 0.8 & -0.8 $\pm$ 0.8 & $<10^{-16}$ & 0.88 \\
2017 12 18 3:19 & 105.644 & 0.163 & 7259 & 3.1 $\pm$ 2.1 & 2.0 $\pm$ 2.1 & 0.82 & 0.86 \\
2017 12 19 6:12 & 106.764 & 0.386 & 16276 & -5.7 $\pm$ 0.9 & 0.4 $\pm$ 0.9 & 3.94 $\times 10^{-7}$ & 0.89 \\
\hline
HD 29615 & & & & & & & \\
\hline
2017 12 13 3:13 & 100.638 & 0.034 & 10063 & -5.0 $\pm$ 3.2 & -1.4 $\pm$ 3.2 & $<10^{-16}$ & 0.79 \\
2017 12 14 1:14 & 101.555 & 0.429 & 6931 & -13.1 $\pm$ 4.6 & 4.2 $\pm$ 4.7 & 8.82 $\times 10^{-8}$ & 0.36 \\
2017 12 14 5:58 & 101.752 & 0.514 & 9573 & -20.8 $\pm$ 3.3 & -0.7 $\pm$ 3.3 & 3.10 $\times 10^{-7}$ & 0.98 \\
2017 12 15 3:52 & 102.665 & 0.907 & 12238 & -23.8 $\pm$ 2.6 & -0.8 $\pm$ 2.6 & $<10^{-16}$ & 0.91 \\
2017 12 16 4:51 & 103.706 & 0.356 & 9819 & 11.5 $\pm$ 3.3 & 0.3 $\pm$ 3.2 & $<10^{-16}$ & 0.97 \\
2017 12 17 6:29 & 104.774 & 0.816 & 10001 & -31.0 $\pm$ 3.2 & -4.2 $\pm$ 3.2 & $<10^{-16}$ & 0.47 \\
2017 12 19 4:26 & 106.688 & 0.641 & 5445 & -13.9 $\pm$ 5.9 & 3.2 $\pm$ 5.9 & 0.018 & 0.80 \\
\hline
HD 35296 & & & & & & & \\
\hline
2017 12 13 4:57 & 100.712 & 0.955 & 15693 & -3.8 $\pm$ 1.5 & 0.2 $\pm$ 1.5 & 5.65 $\times 10^{-6}$ & 0.87 \\
2017 12 14 3:47 & 101.663 & 0.227 & 12256 & -4.3 $\pm$ 1.9 & 2.0 $\pm$ 1.9 & 1.41 $\times 10^{-4}$ & 0.95 \\
2017 12 15 5:39 & 102.741 & 0.536 & 16769 & 2.4 $\pm$ 1.4 & -1.8 $\pm$ 1.5 & 4.81 $\times 10^{-3}$ & 0.91 \\
2017 12 16 3:03 & 103.633 & 0.791 & 16040 & 3.8 $\pm$ 1.4 & -1.8 $\pm$ 1.4 & 1.77 $\times 10^{-3}$ & 0.85 \\
2017 12 16 5:48 & 103.747 & 0.824 & 17868 & 2.4 $\pm$ 1.3 & -2.1 $\pm$ 1.3 & 0.050 & 0.91 \\
2017 12 17 3:12 & 104.639 & 0.079 & 16668 & -4.5 $\pm$ 1.4 & -1.8 $\pm$ 1.4 & $<10^{-16}$ & 0.94 \\
2017 12 18 2:47 & 105.622 & 0.360 & 5235 & -3.5 $\pm$ 4.4 & -3.9 $\pm$ 4.4 & 0.94 & 0.97 \\ 
2017 12 19 5:31 & 106.735 & 0.679 & 8908 & 4.1 $\pm$ 2.6 & 2.0 $\pm$ 2.6 & 9.52 $\times 10^{-5}$ & 0.95 \\
\hline
V1358 Ori & & & & & & & \\
\hline
2017 12 13 4:21 & 100.686 & 0.807 & 9163 & -16.6 $\pm$ 8.9 & -12.9 $\pm$ 9.0 & 0.48 & 0.99 \\
2017 12 13 6:47 & 100.788 & 0.882 & 7049 & -33.3 $\pm$ 11.5 & 12.9 $\pm$ 11.5 & 0.35 & 0.91 \\
2017 12 14 4:51 & 101.707 & 0.559 & 6139 & 29.6 $\pm$ 13.1 & 5.7 $\pm$ 13.1 & 0.70 & 0.98 \\
2017 12 15 4:59 & 102.713 & 0.300 & 9952 & -8.8 $\pm$ 8.0 & -6.6 $\pm$ 8.0 & 0.45 & 0.96 \\
2017 12 16 3:45 & 103.661 & 0.999 & 8278 & -20.8 $\pm$ 9.9 & 11.7 $\pm$ 9.9 & 1.53 $\times 10^{-3}$ & 0.89 \\
2017 12 17 3:53 & 104.667 & 0.740 & 9674 & -16.4 $\pm$ 8.4 & -7.4 $\pm$ 8.4 & 0.078 & 0.99 \\ 
2017 12 19 6:55 & 106.793 & 0.307 & 5682 & 1.5 $\pm$ 14.9 & -15.5 $\pm$ 14.1 & 0.56 & 0.97 \\
\hline
\end{tabular}
\tablefoot{$\phi_{\mathrm{rot}}$ is the rotational phase. $S/N$ is the signal-to-noise ratio of the Stokes LSD V profile. $\langle B_z \rangle$ is shown both for the Stokes V profile and the null profile.}
\end{table*}

\begin{table}
\centering
\caption{\label{obs2}Summary of the observations.}
\begin{tabular}{c c c c c}
\hline\hline
Star & $\langle S/N\rangle$ & $n_{\mathrm{LSD}}$ & $n_{\phi}$ & $f_{\phi}$ \\
\hline
BE Cet & 17833 & 5740 & 7 & 0.700 \\
$\chi^1$ Ori & 17154 & 4790 & 8 & 0.543 \\
HD 29615 & 9153 & 5715 & 7 & 0.649 \\
HD 35296 & 13680 & 4782 & 8 & 0.733 \\
V1358 Ori & 7991 & 3730 & 7 & 0.549 \\
V1358 Ori (2013) & 8800 & 3085 & 4 & 0.400 \\
\hline
\end{tabular}
\tablefoot{$\langle S/N \rangle$ is the mean signal-to-noise ratio of the Stokes LSD V profiles, $n_{\mathrm{LSD}}$ is the number of lines included in the LSD profile, $n_{\phi}$ is the number of observed phases, and $f_{\phi}$ is the phase coverage.}
\end{table}

\section{Zeeman-Doppler imaging}

The idea of ZDI is to model polarised spectral line profiles formed by the distribution of magnetic fields on the surface of a rotating star. The Stokes V profiles, observed at as many evenly distributed rotational phases as possible, are inverted into the magnetic field that would best reproduce the observed line profiles. The observations should not cover more than several rotation periods, in order for the magnetic field and spot configurations not to evolve significantly during the observations. For the more rapid rotators, with sufficiently broad spectral lines, a brightness map is also produced from the Stokes I profiles. A practical requirement for this is around $v\sin i \gtrsim 15$ km s$^{-1}$, although it depends on the spectral resolution of the instrument. Each phase is weighted as proportional to the square of its signal-to-noise ratio ($S/N$), measured from the LSD V profile; these values are shown in Table \ref{obs}.

We used the code inversLSD, developed by \cite{kochukhov2014_LSD}, to invert the ZDI maps. 
The stellar surface is divided into 1876 elements, and the radial, meridional, and azimuthal magnetic field components and the brightness are calculated for each. 
Instead of directly modelling the magnetic field components, the magnetic field is represented with spherical harmonic expansion coefficients $\alpha_{\ell,m}$, $\beta_{\ell,m}$, $\gamma_{\ell,m}$, where $\alpha$ represents the radial poloidal component, $\beta$ the horizontal poloidal component, and $\gamma$ the horizontal toroidal component of the field. The $\ell$ is the angular degree, and $m$ is the azimuthal order of the mode. 
Higher-order harmonics represent the magnetic field on smaller scales. This allows a straightforward calculation of the magnetic energy stored in the magnetic field on different size scales. For the number of spherical harmonic expansions,
\cite{hackman16} found that increasing the maximum angular degree, $\ell_{\mathrm{max}}$, to more than 5 did not significantly improve the fit to the Stokes V profiles. Also, \cite{lehtinen2019arXiv} noted that, although they used $\ell_{\mathrm{max}} = 20$, most of the magnetic energy was found in harmonics below $\ell \leq 6$. These results were, however, derived for stars with higher $v \sin i$ than some of the stars in our sample (notably BE Cet and $\chi^1$ Ori), so they are not directly applicable to these slower rotators. 
We have limited the solution to $\ell_{\mathrm{max}} = 10$ harmonics for all our stars, which gives a sufficient resolution to study the large-scale properties of the magnetic field.

The magnetic field and brightness inversions are performed simultaneously so that brightness inhomogeneities are taken into account in the magnetic field inversion. For the magnetic field, high-order modes are suppressed by a regularisation function. For the brightness inversion, Tikhonov regularisation is used to limit the surface gradient of the solution. A smoother variation is thus preferred over steep variations. 
A more thorough description of the regularisation can be found in \cite{rosen16}.

The inversLSD code includes a parameter for the differential rotation ($\alpha = \Delta\Omega / \Omega_{\mathrm{eq}}$, where $\Delta\Omega$ is the rotational shear between the equator and the poles, and $\Omega_{\mathrm{eq}}$ is the angular velocity at the equator), which, if significant, can affect the result. Commonly, $\alpha$ is assumed to be reduced as a function of rotation. There is theoretical \citep{kitchatinov1999} and numerical \citep{viviani2018} evidence for this. Generally, observations agree with these results \citep[e.g.][]{reiners03,reinhold2015,lehtinen16}, although the absolute shear, $\Delta\Omega$, was found by \cite{reinhold2015} to slightly increase towards shorter rotation periods. However, some inconsistent results have also been reported; for example, V889 Her is a young solar analogue with a rotation period of about 1.3 days, and some studies have found it to have very high differential rotation \citep{marsden06,jeffers08}.

We have assumed the stars to rotate as solid bodies. For those stars with published estimates for the differential rotation, we have also made alternative maps with those values to check how much it might affect the large-scale structure of the magnetic field.

The photometric rotation period, $P_{\mathrm{phot}}$, is measured from spots, which can be at high latitudes. Since the rotation period, $P_{\mathrm{rot}}$, used in the model (listed for all stars in Table \ref{param}) is imposed at the equator, the inclusion of solar-like differential rotation (faster rotation at the equator and slower at the poles) requires a shorter rotation period than a model with solid body rotation. In these cases, we searched for the best $P_{\mathrm{rot}}$ by minimising the deviation of the model and observations for the Stokes V data. We used the reported differential rotation and tried smaller values of $P_{\mathrm{rot}}$ with a step of 0.01 d for the other stars, and 0.005 d for V1358 Ori (the fastest rotator), until a $\chi^2$ minimum was found. This was then chosen as $P_{\mathrm{rot}}(\alpha \neq 0)$ for the case with differential rotation.

\section{Adopted stellar parameters}

This section presents the objects of this study and their respective adopted stellar parameters, which are summarised in Table \ref{param}.
There are some problems with these parameters. The stellar radius, $R$, can be estimated from the used parameters with the equation

\begin{equation} \label{radius}
R = \frac{P_{\mathrm{rot}} v\sin i}{2\pi \sin i},
\end{equation}

\noindent where $v \sin i$ is the line-of-sight component of the rotational velocity at the equator, and $i$ is the inclination. When comparing this radius to values found from the literature, there is good agreement only with $\chi^1$ Ori and HD 29615 (see Table \ref{param}). Furthermore, the combination of parameters found from the literature is mathematically impossible for BE Cet and V1358 Ori. 
Besides the possibility that there are errors in the values for the radius, the inclination is probably the most uncertain parameter. Since ZDI is not very sensitive to small errors in the inclination, we chose to use the parameters found from the literature despite it being clear that there are some problems with them.

In Table \ref{param} we have also listed the Rossby number for the stars, $\mathrm{Ro} = P_{\mathrm{rot}} / \tau_{\mathrm{conv}}$, where $\tau_{\mathrm{conv}}$ is the convective turnover time. The Rossby number, which anti-correlates with magnetic activity, is a better parameter to indicate the level of activity than the rotation period alone \citep{lehtinen2020}. Ro is inversely proportional to the Coriolis number, Co, which is commonly used as an indicator for rotation in numerical simulations.
The convective turnover times are calculated from the B-V colour index, taken from the Hipparcos Catalogue \citep{ESA1997_HIP}, using the empirical equation presented for main-sequence stars by \cite{noyes84}.

\subsection{BE Cet}

\object{BE Cet} (HD 1835, HIP 1803) is a star of spectral type G2.5V \citep{keenan1989} with an effective temperature $T_{\mathrm{eff}} = 5837$ K \citep{valenti2005}. It has a radius $R$ = 1.00 $R_\odot$ \citep{takeda2007}, an age of around 600 Myr \cite[see][]{rosen16}, and a chromospheric activity index $\log R'_{\mathrm{HK}} = -4.454$ \citep{lehtinen2020}. It is the slowest rotator in our sample, with $P_{\mathrm{rot}} = 7.676$ d \citep{lehtinen2020} and $v\sin i = 7.0$ km s$^{-1}$ \citep{valenti2005}, and also has the highest Rossby number, $\rm{Ro} = 0.61$. A 9.1 yr chromospheric cycle was reported by \cite{baliunas1995}. A previous ZDI map from 2013 was published by \cite{rosen16}.

Our observations of BE Cet are phased with the ephemeris

\begin{equation}
\mathrm{HJD}_{\phi=0} = 2456545.580 + 7.676 \times E.
\end{equation}

\subsection{$\chi$1 Ori}

The star \object{$\chi^1$ Ori} (HD 39587, HIP 27913) has a spectral type of G0V \citep{keenan1989} with $T_{\mathrm{eff}} = 5882$ K \citep{valenti2005}. It has a radius $R$ = 1.05 $R_\odot$ \citep{takeda2007} and an age around 300 Myr \citep[see][]{rosen16}. It has a rotation period $P_{\mathrm{rot}} = 5.022$ d \citep{lehtinen2020}, $v\sin i = 9.8$ km s$^{-1}$ \citep{valenti2005}, $\rm{Ro} = 0.57$, and $\log R'_{\mathrm{HK}} = -4.456$ \citep{lehtinen2020}. It is classified as variable by \cite{baliunas1995}, meaning that it shows variable levels of chromospheric emission but lacks clear cyclicity. Previous ZDI maps from 2007, 2008, 2010, and 2011 were published by \cite{rosen16}. In those maps, the magnetic polarity was reversed twice over 4.8 years, indicating a possible rapid magnetic cycle.

Our observations of $\chi^1$ Ori are phased with the ephemeris

\begin{equation}
\mathrm{HJD}_{\phi=0} = 2454127.402 + 5.022 \times E.
\end{equation}

\subsection{HD 29615}

\object{HD 29615} (HIP 21632) is a star of spectral type G3V \citep{torres2006} with $T_{\mathrm{eff}} = 5820$ K \citep{waite15}. It has a radius $R$ = 0.96 $R_\odot$ and rotation period $P_{\mathrm{rot}} = 2.32$ d \citep{vidotto2014}, $v\sin i = 18.5$ km s$^{-1}$ \citep{hackman16}, and $\rm{Ro} = 0.24$, making it the second most active star in our sample. Previous ZDI maps of HD 29615 were published by \cite{waite15} and \cite{hackman16}. When comparing those maps, \cite{hackman16} noted a polarity reversal in the magnetic field between 2009 and 2013, when their respective data were obtained.

\cite{waite15} also derived values for the differential rotation of HD 29615. They gained very different values for the Stokes I and V data: $\Delta\Omega = 0.07^{+0.10}_{-0.03}$ rad d$^{-1}$ (Stokes I) and $\Delta\Omega = 0.48^{+0.11}_{-0.12}$ rad d$^{-1}$ (Stokes V).

Our observations of HD 29615 are phased with the ephemeris

\begin{equation}
\mathrm{HJD}_{\phi=0} = 2449730.0 + 2.32 \times E.
\end{equation}

\subsection{HD 35296}

\object{HD 35296} (111 Tau, HIP 25278) is a star of spectral type F8V \citep{montes2001,gray2003} and $T_{\mathrm{eff}} = 6170$ K \citep{casagrande2011}. It has a rotation period $P_{\mathrm{rot}} = 3.493$ d \citep{lehtinen2020}, $v\sin i = 15.9$ km s$^{-1}$ \citep{waite15}, and $\rm{Ro} = 0.57$, a radius $R$ = 1.10 $R_\odot$ \citep{zuckerman2011}, and $\log R'_{\mathrm{HK}} = -4.426$ \citep{lehtinen2020}. Age estimates of HD 35296 vary considerably; \cite{barry1988} estimated from the chromospheric activity-age relation a young age of 20 Myr, whereas \cite{holmberg2009} reported an age of 3.3 Gyr based on evolutionary models, and \cite{chen2001} estimated from the Li-abundance an age of 7.5 Gyr. \cite{li1998} listed it as a probable member of the Taurus-Auriga star forming region, which would again imply that it is a young star. \cite{waite15} argued that the star is young, between 20-50 Myr, based on theoretical isochrones. The star is included in the Mount Wilson sample, but no cycle, or a cycle too long to be defined, was reported by \cite{baliunas1995}. Two earlier ZDI maps from 2007 and 2008 were published by \cite{waite15}. In both maps, the radial magnetic field was dominated by small-scale structures with mixed polarities. \cite{waite15} also derived a value for the differential rotation of HD 35296 from Stokes V data: $\Delta\Omega = 0.22^{+0.04}_{-0.02}$ rad d$^{-1}$.

Our observations of HD 35296 are phased with the ephemeris

\begin{equation}
\mathrm{HJD}_{\phi=0} = 2454496.094 + 3.493 \times E.
\end{equation}

\subsection{V1358 Ori}

\object{V1358 Ori} (HD 43989, HIP 30030) is a star of spectral type F9V \citep{montes2001} and has an effective temperature $T_{\mathrm{eff}} = 6032$ K \citep{mcdonald2012}. Its radius is slightly larger than the Sun's, with published values of $R = 1.08$ $R_\odot$ \citep{zuckerman2011} and $R = 1.05$ $R_\odot$ \citep{vican2014}. A previous ZDI map from 2013 was published by \cite{hackman16}. They, however, used an incorrect rotation period, $P_{\mathrm{rot}} = 1.16$ d, which was photometrically derived by \cite{cutispoto2003}. Using a much more extensive photometric dataset, \cite{kriskovics2019} derived the rotation period $P_{\mathrm{rot}} = 1.3571$ d, which we have adopted. We thus also re-analysed the 2013 data from \cite{hackman16} for this star using the new rotation period. As a fast rotator, with $v\sin i = 41.3$ km s$^{-1}$, V1358 Ori is the most active star in our sample, with $\rm{Ro} = 0.16$.

\cite{kriskovics2019} also applied Doppler imaging to the star. Their two maps, derived from independent subsets of the same observations from 2013, are both dominated by a cool polar spot, but lower latitude spots and hot surface features are also present in both maps. They conclude that these are probably real features since they are visible in both independent subsets of their data. They also found a 4.5 yr photometric cycle and a differential rotation coefficient $\alpha = 0.016$, corresponding to an absolute shear $\Delta\Omega = 0.07$ rad d$^{-1}$.

Our observations of V1358 Ori are phased with the ephemeris

\begin{equation}
\mathrm{HJD}_{\phi=0} = 2449681.5 + 1.3571 \times E.
\end{equation}

\begin{table*}
\centering
\caption{Stellar parameters and parameters used in the ZDI inversion.}
\label{param}
\begin{tabular}{c c c c c c}
\hline\hline
Parameter & BE Cet & $\chi^1$ Ori & HD 29615 & HD 35296 & V1358 Ori \\
\hline
$P_{\mathrm{rot}}$ [d] & 7.676$^a$ & 5.022$^a$ & 2.32$^b$ & 3.493$^a$ & 1.3571$^c$ \\
$\tau_{\mathrm{conv}}$ [d] & 12.51 & 8.76 & 9.72 & 6.12 & 8.37 \\
Ro & 0.61 & 0.57 & 0.24 & 0.57 & 0.16 \\
$v\sin i$ [km s$^{-1}$] & 7.0$^d$ & 9.8$^d$ & 18.5$^e$ & 15.9$^f$ & 41.3$^e$ \\
Inclination [$^\circ$] & 65$^g$ & 65$^g$ & 62$^e$ & 65$^f$ & 59$^e$ \\
$R$ (from Eq. \ref{radius}) [$R_\odot$] & 1.17 & 1.07 & 0.96 & 1.21 & 1.29 \\
$R$ (literature) [$R_\odot$] & 1.00$^h$ & 1.05$^h$ & 0.96$^b$ & 1.10$^f$ & 1.05$^i$/1.08$^j$ \\
$T_{\mathrm{eff}}$ [K] & 5837$^d$ & 5882$^d$ & 5820$^f$ & 6170$^k$ & 6032$^l$ \\
Effective Land\'{e} $g$ factor & 1.216 & 1.217 & 1.213 & 1.218 & 1.218 \\
\hline
\end{tabular}
\tablefoot{References: $^a$\cite{lehtinen2020}, $^b$\cite{vidotto2014}, $^c$\cite{kriskovics2019}, $^d$\cite{valenti2005}, $^e$\cite{hackman16}, $^f$\cite{waite15}, $^g$\cite{rosen16}, $^h$\cite{takeda2007}, $^i$\cite{vican2014}, $^j$\cite{zuckerman2011}, $^k$\cite{casagrande2011}, $^l$\cite{mcdonald2012}.}
\end{table*}

\section{Results} \label{results}

The ZDI maps for each star, for the radial, meridional, and azimuthal components of the magnetic field, are presented in Figs. \ref{becet}--\ref{v1358ori}, along with the brightness map for those stars with sufficiently large $v\sin i$ (> 15 km s$^{-1}$) to allow its production. The spherical harmonic decomposition of the magnetic field allows us to study how the magnetic energy is distributed between the poloidal and toroidal field and on different angular scales. 
The maximum and average strengths of the total magnetic field are summarised for all stars in Table \ref{feat}, along with the amounts of poloidal versus toroidal and axisymmetric 
versus non-axisymmetric 
components. For axisymmetry, we have followed the definition of \cite{fares2009_tauboo}, where spherical harmonic modes $|m| < \ell / 2$ are defined as axisymmetric and modes $|m| \geq \ell / 2$ as non-axisymmetric, as has been done in many previous studies \citep[e.g.][]{hackman16,rosen16,lehtinen2019arXiv}. We followed this definition for the sake of consistency, but we also calculated the fractions of axisymmetric versus non-axisymmetric components using an alternative, perhaps mathematically more intuitive, definition of $m=0$ for the axisymmetric component, which is also used in some studies \citep[e.g.][]{waite15}. These fractions are shown in Appendix \ref{append4}.

The information in Table \ref{feat} is also shown in Fig. \ref{deconfusogram} in a \textquoteleft deconfusogram' (a modification of the standard confusogram), introduced by \cite{pineda2020}. 
One additional data point of BE Cet from 2013 and four data points of $\chi^1$ Ori from 2007, 2008, 2010, and 2011 were added from the results of \cite{rosen16}, along with one data point of HD 29615 from 2013 from \cite{hackman16}.
\footnote{We used the values reported in those papers. Additionally, we used 85 \% and 12 \% for the fraction of poloidal and axisymmetric magnetic field, respectively, for BE Cet; used 56 \% and 60 \% (2007), 26 \% and 76 \% (2008), 22 \% and 78 \% (2010), and 35 \% and 67 \% (2011) for the same parameters for $\chi^1$ Ori; and used $\langle B\rangle = 84$ G for HD 29615. These values were provided by the authors of the original papers.} For the deconfusogram, we used a modified version of the freely available plotting code from \cite{pineda2020}.

Additional parameters of the ZDI maps, the fractions of axisymmetric and non-axisymmetric magnetic fields using the alternative definition $m = 0$ for axisymmetry, and the mean absolute value of the magnetic field for the radial, meridional, and azimuthal components are tabulated in Appendix \ref{append4}.

We calculated Pearson correlation coefficients, $r$, for the correlation between the brightness and the absolute values of the different components of the magnetic field for the stars with a brightness map. They, as well as the deviation between model and observation $\sigma$ for the Stokes V profiles, are also included in Table \ref{feat}. The distribution of the total magnetic energy, in the poloidal and toroidal component combined, over the spherical harmonics $\ell$ for all our maps is shown in Fig. \ref{el} and listed in Appendix \ref{append3}.

It should be remembered that ZDI is more effective in capturing the magnetic field on large scales, and much of the small-scale field, as much as 90--99 \%, will go undetected \citep{kochukhov2020}. This means that the global properties of the maps are probably quite trustworthy, but the smaller details are uncertain.

\subsection{BE Cet}

\begin{figure*}
   \centering
   \includegraphics[bb=30 200 800 340,width=\textwidth,angle=180]{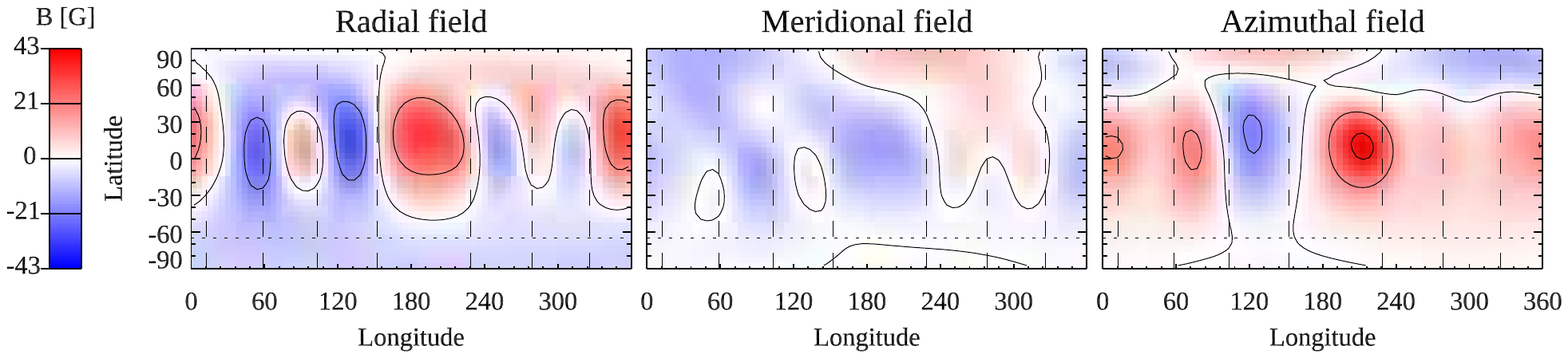}
   \caption{ZDI map for BE Cet. The vertical lines indicate the phases of individual spectra. The horizontal line marks the inclination of the star. For the parts below it, around the invisible pole, there is no information in the data.}
   \label{becet}
\end{figure*}

Our magnetic field maps of BE Cet are presented in Fig. \ref{becet}. Due to the line profiles being fairly narrow because of the star's comparably slow rotation, we did not produce a brightness map.

Our map shows a strong poloidal (73 \%) and non-axisymmetric (75 \%)  component, more so than with any of the other stars. This was also the case in the observations from 2013, reported by \cite{rosen16}, and the magnetic field strength is very similar to that in their map. In Fig. \ref{el}, BE Cet also stands out as the star with the most total magnetic energy in higher-order spherical harmonics ($2 \leq \ell \leq 4$). 
In \cite{rosen16}, BE Cet was also among the stars with more energy in these harmonics. No signs of polarity reversals are detected.

\subsection{$\chi$1 Ori}

\begin{figure*}
   \centering
   \includegraphics[bb=30 200 800 340,width=\textwidth,angle=180]{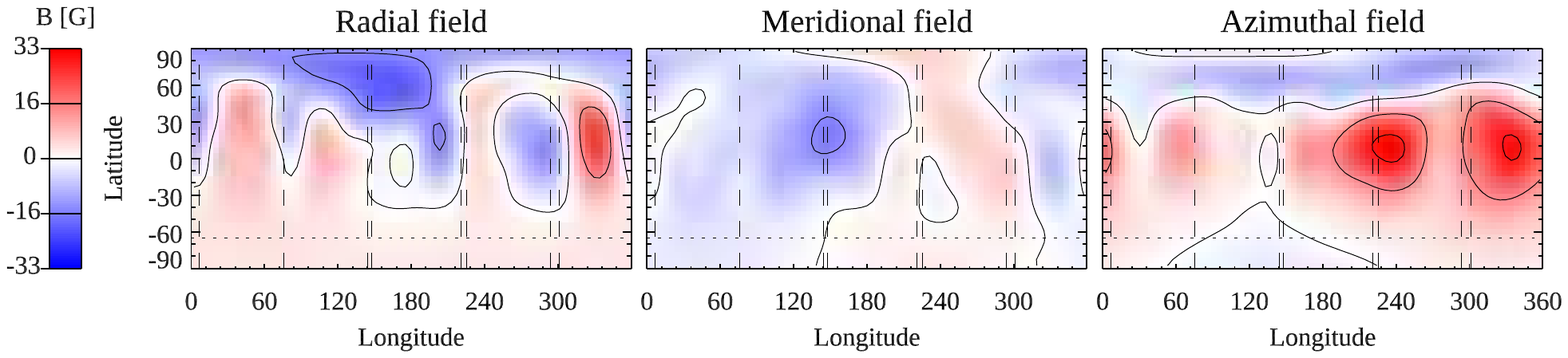}
   \caption{Same as Fig. \ref{becet}, but for $\chi^1$ Ori.}
   \label{chi1ori}
\end{figure*}

The magnetic field maps of $\chi^1$ Ori are presented in Fig. \ref{chi1ori}. The $v \sin i$  of  $\chi^1$ Ori is too low to produce a brightness map. The poloidal and toroidal components (48 \% and 52 \%) as well as the axisymmetric and non-axisymmetric components (53 \% and 47 \%) of the magnetic field are very similar in strength. In the results of \cite{rosen16}, the field had a stronger axisymmetric component and there was much variation between the different datasets for the amount of poloidal and toroidal field.\ Here, $\ell = 1$ is the dominating mode with the most magnetic energy.

In our observations from December 2017, the polarity of the radial field is oriented in the same way as in the last dataset of \cite{rosen16}, which was from October-November 2011. If the rapid polarity switches of the star, seen in the results of \cite{rosen16}, have continued, it may have switched polarity twice between their October-November 2011 and our December 2017 observations, so the magnetic field would again have the same polarity. This would indicate a rapid magnetic cycle of only a few years, but more frequent mapping would be needed to verify this. We can estimate the cycle period of $\chi^1$ Ori by assuming a perfectly periodic cycle that always has the same interval between polarity reversals. If we assume that two polarity reversals occurred between October-November 2011 and December 2017 and require that the polarity of the radial magnetic field agree with the ZDI maps from January-February 2007, January-February 2008, September-October 2010, and October-November 2011 from \cite{rosen16}, this would allow a magnetic cycle period of between 3.8 and 7.2 years, with polarity reversals with an interval of between 1.9 to 3.6 years. Since actual cycles are not perfectly periodic, this should be seen only as a crude estimate of when new polarity reversals could be expected. 
Nevertheless, this star seems to be an interesting target for future observations, presenting a possibility to characterise a magnetic cycle in a reasonably short time, similar to what has been done with $\tau$ Boo \citep{mengel2016_tauboo,jeffers2018_tauboo}, 61 Cyg A \citep{boro_saikia2016,boro_saikia2018}, and HD 75332 \citep{brown2021_hd75332}.

\subsection{HD 29615}

\begin{figure*}
   \centering
   \includegraphics[bb=30 200 800 340,width=\textwidth,angle=180]{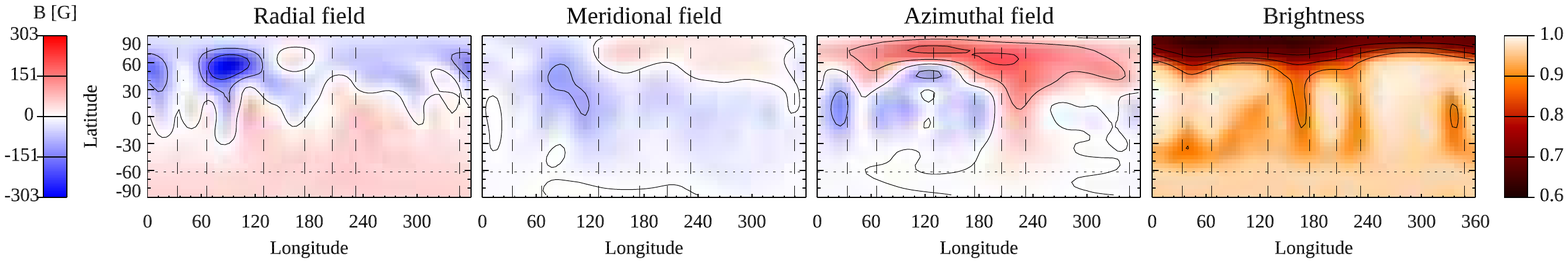}
   \caption{Same as Fig. \ref{becet}, but for HD 29615, with the addition of the brightness map.}
   \label{hd29615}
\end{figure*}

The ZDI maps for HD 29615 are presented in Fig. \ref{hd29615}. In our maps, the orientation of the magnetic field is similar to as in the 2013 observations of \cite{hackman16}. Our radial field does display a more complex configuration, notably a strong concentration of radial field in a small high latitude region. It is also notable, though, that there is a large gap in our phase coverage, and the featureless regions on the stellar surface coincide with these poorly observed phases. Some features may thus have been missed at those longitudes.

This star has the strongest magnetic field of the stars in our sample. 
The brightness map is dominated by a large polar spot, as previously reported by both \cite{waite15} and \cite{hackman16}. No strong correlations are found between the brightness and the magnetic field components, as was also reported by \cite{hackman16}. The magnetic field is slightly dominated by the poloidal component (65 \%), similar to what was reported by \cite{hackman16} (62.8 \%), but to a lesser degree than reported by \cite{waite15} (75 \%). The axisymmetric component (61 \%) is larger than the non-axisymmetric one, similar to what was reported by \cite{waite15} (66 \%) but to a lesser extent than with \cite{hackman16} (87.6 \%). We note, however, that \cite{waite15} used the definition $m=0$ for the axisymmetric component. With this definition we would have a very even distribution between the axisymmetric component (51 \%) and the non-axisymmetric one. Most of the magnetic energy is concentrated in the $\ell = 1$ mode.

We also produced an alternative ZDI map of HD 29615, taking into account the high value for differential rotation reported for the Stokes V data by \cite{waite15}. We note, though, that for the Stokes I data, they found a considerably lower value, which we deemed negligible. With their value for differential rotation, which translates to $\alpha = 0.177$, we found the best period to be $P_{\mathrm{rot}} = 2.23$ d. 
The alternative ZDI map, constructed with these parameters, is shown in Appendix \ref{hd29615_difrot}. On large scales, the map is essentially similar to the one without differential rotation and thus does not significantly change the results.

\subsection{HD 35296}

\begin{figure*}
   \centering
   \includegraphics[bb=30 200 800 340,width=\textwidth,angle=180]{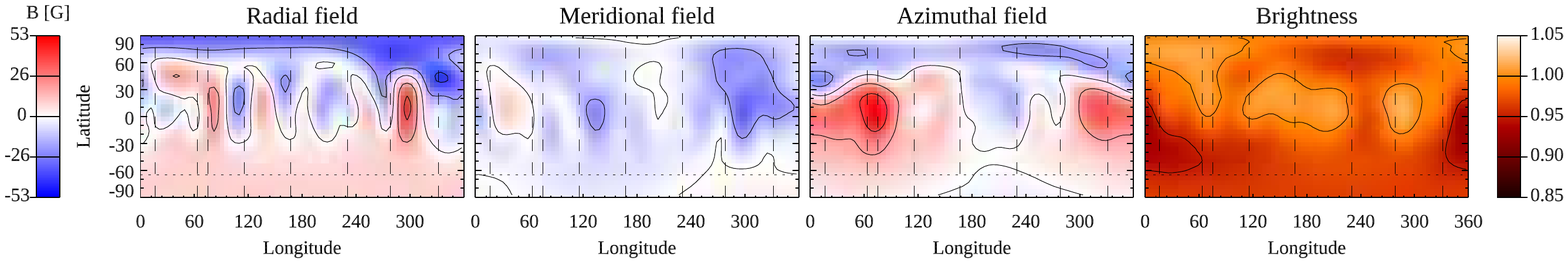}
   \caption{Same as Fig. \ref{hd29615}, but for HD 35296.}
   \label{hd35296}
\end{figure*}

Our maps for HD 35296 are shown in Fig. \ref{hd35296}. The magnetic field strength is close to that in the maps of \cite{waite15}. In our radial field map, there are features of mixed polarities, but the large-scale orientation of the magnetic field is still much clearer than in the maps of \cite{waite15}. There might have been a polarity reversal between 2007 and 2017 since the radial magnetic field orientation in the polar regions is positive in the map for 2007 in \cite{waite15} and then very mixed in their map for 2008, while in our map for 2017 it is clearly negative. The lack of clear large-scale structures in their maps, however, makes the situation quite unclear. In our map, the poloidal component is slightly larger than the toroidal (63 \%). In the maps of \cite{waite15}, the poloidal component was first 82 \% in 2007 and then 50 \% in 2008. In our map, the axisymmetric and non-axisymmetric components are of similar strength (52 \% and 48 \%). This is similar to the 2008 data of \cite{waite15} (45 \% axisymmetric), but in their 2007 data the field was dominated by the non-axisymmetric component (19 \% axisymmetric). Again, using the same definition as \cite{waite15} for the axisymmetric component, we would get a less axisymmetric field (44 \%) for our map.

This is also the only one of the stars with sufficiently rapid rotation for the generation of a brightness map that does not show any cool polar spot. There is one large spot feature at lower latitudes, but with a fairly moderate contrast. At higher latitudes there are some possible brighter features. Their reliability, however, is uncertain since bright features are a known artefact in Doppler images \citep[e.g.][]{berdyugina1998}. It is, however, notable, that the $v\sin i$ of this star is quite low, close to the limit where the spectral features cannot be resolved properly and a meaningful brightness map can no longer be constructed. For example, \cite{waite15} did not publish a brightness map, although their observations were done with a lower spectral resolution (around $R \approx$ 65 000 for NARVAL vs. our $R \approx$ 110 000 for HARPS). No strong correlations are found between the brightness and the magnetic field components. Most of the magnetic energy is concentrated in the $\ell = 1$ mode.

We also repeated our analysis, taking into account the differential rotation value, $\alpha = 0.122$, of \cite{waite15}. Since the deviation between the model and observations was much larger when including the differential rotation than with solid body rotation\footnote{The deviation between model and observations for Stokes V data was $\sigma = 6.90 \times 10^{-5}$ with $\alpha = 0$ (and $P_{\mathrm{rot}} = 3.493$ d) vs. $\sigma = 7.12 \times 10^{-5}$ with $\alpha = 0.122$ and $P_{\mathrm{rot}} = 3.49$ d, and $\sigma = 7.43 \times 10^{-5}$ with $\alpha = 0.122$ and $P_{\mathrm{rot}} = 3.40$ d.}, we deemed the solid body rotation model to be valid for HD 35296. Furthermore, the deviation increased with decreasing $P_{\mathrm{rot}}$. With solar-like differential rotation it would be expected to decrease since the rotation period for the ZDI inversion is measured at the equator, which would be rotating more rapidly with the inclusion of differential rotation.

\subsection{V1358 Ori}

\begin{figure*}
   \centering
   \includegraphics[bb=30 200 800 340,width=\textwidth,angle=180]{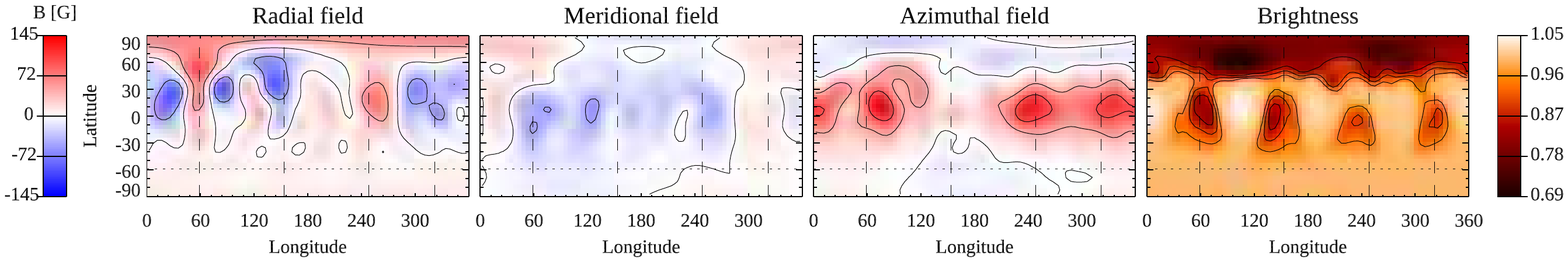}
   \includegraphics[bb=30 200 800 340,width=\textwidth,angle=180]{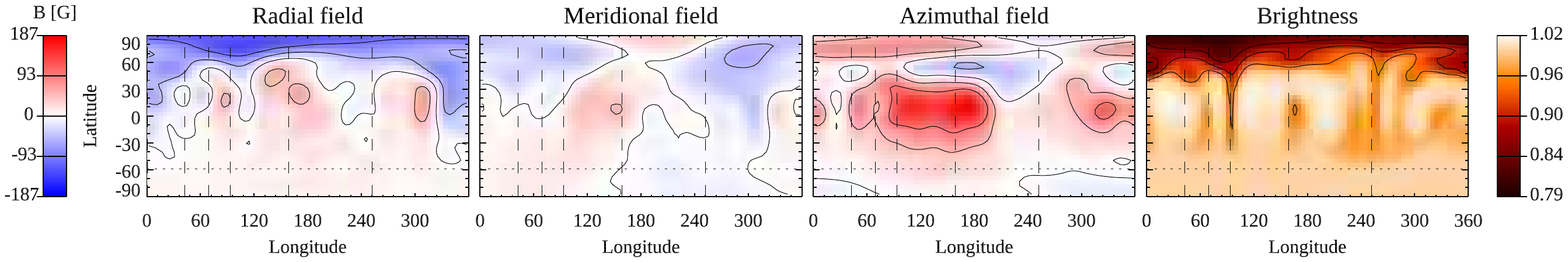}
   \caption{Same as Fig. \ref{hd29615}, but for V1358 Ori, for 2013 (upper map) and 2017 (lower map).}
   \label{v1358ori}
\end{figure*}

The ZDI maps for V1358 Ori, for both our new 2017 observations and the older re-analysed 2013 observations, are shown in Fig. \ref{v1358ori}. The most interesting feature in the maps is the polarity reversal, which occurred between 2013 and 2017. No long-term photometric or chromospheric data are available for V1358 Ori, so this polarity reversal cannot be related to a cycle. The field strength increased slightly from 2013 to 2017.

In the brightness map published by \cite{hackman16}, there is no evident polar spot. However, in our map of the same data, the pole is covered by a dark feature, as it is in our map from the 2017 data. The polar spot is also confirmed in the results of \cite{kriskovics2019} from  data from 2013, so it seems that the spot activity of V1358 is mostly concentrated in polar regions. Some lower latitude features are also visible in our maps, but at least in the 2013 map they are clearly concentrated closely around the observed phases. The bad phase coverage of the observations can lead to the inversion code preferring solutions with lower latitude spots at the observed phases, so these might be artefacts. 
For instance, in their study of 41 Doppler maps of HD 199178, a star with a dominant polar spot, \cite{hackman2019} note that all the maps with low latitude spot activity can be explained with a poor phase coverage.

The 2017 map of V1358 Ori is the only one in our whole sample to show a fairly strong correlation between the brightness and the radial magnetic field ($r_{\mathrm{rad}} = 0.55$). In the 2013 map this correlation is not seen ($r_{\mathrm{rad}} = 0.16$). The result for 2013 is not very reliable due to the poor phase coverage, but with more observations of good quality it could be interesting to see if the correlations change systematically over the activity cycle. If so, it could tell us if the magnetic field is dominated by large or small features during different phases of the cycle. The correlation should be better for a large-scale field and worse for a field dominated by small-scale features that cancel one another out and go undetected.

The amounts of poloidal versus toroidal field and axisymmetric versus non-axisymmetric field are very similar in the 2013 and 2017 maps. V1358 is the star with the largest toroidal component (59 \% in both maps). In both maps, $\ell=1$ is the dominating mode. However, we again note that the values for the map from 2013 in particular are not very reliable due to the poor phase coverage. Also, due to the high noise level, the results for V1358 Ori might be subject to larger uncertainties than for the other stars.

Due to the results of \cite{kriskovics2019}, we also repeated our analysis of V1358 Ori using their value for the differential rotation: $\alpha = 0.016$. Here we found the best period to be $P_{\mathrm{rot}} = 1.335$ d. An alternative ZDI map, with these parameters, is shown in Appendix \ref{v1358ori_difrot}. The large-scale structure of the magnetic field is the same as without the inclusion of differential rotation, which would thus not affect our conclusion of the polarity reversal.

\begin{table*}
\centering
\caption{Summary of the magnetic field properties.}
\label{feat}
\begin{tabular}{c c c c c c c c c c c}
\hline\hline
Star & $B_{\mathrm{max}}$ [G] & $\langle B\rangle$ [G] & Pol. [\%] & Tor. [\%] & Axis. [\%] & Non-axis. [\%] & $r_{\mathrm{rad}}$ & $r_{\mathrm{mer}}$ & $r_{\mathrm{azi}}$ & $\sigma$ [$10^{-5}$] \\
\hline
BE Cet & 55 & 16 & 73 & 27 & 25 & 75 & \ldots & \ldots & \ldots & 4.70 \\
$\chi^1$ Ori & 41 & 13 & 48 & 52 & 53 & 47 & \ldots & \ldots & \ldots & 5.47 \\
HD 29615 & 329 & 80 & 65 & 35 & 61 & 39 & 0.21 & 0.03 & 0.34 & 9.23 \\
HD 35296 & 58 & 21 & 63 & 37 & 52 & 48 & 0.24 & 0.26 & 0.09 & 6.90 \\
V1358 Ori (2013) & 154 & 50 & 41 & 59 & 62 & 38 & 0.16 & 0.01 & 0.15 & 12.5 \\
V1358 Ori (2017) & 195 & 58 & 41 & 59 & 63 & 37 & 0.55 & 0.26 & 0.07 & 11.5 \\ 
\hline
\end{tabular}
\tablefoot{The columns show: the name of the star; the maximum value of the magnetic field, $B_{\mathrm{max}}$; the average value of the magnetic field, $\langle B\rangle$; the fractions of poloidal, toroidal, axisymmetric, and non-axisymmetric fields; the correlation between the brightness map and the radial, meridional, and azimuthal magnetic field maps; and the deviation between model and observations for the Stokes V LSD profiles, $\sigma$.}
\end{table*}

\begin{figure}
   \centering
   \includegraphics[width=9cm]{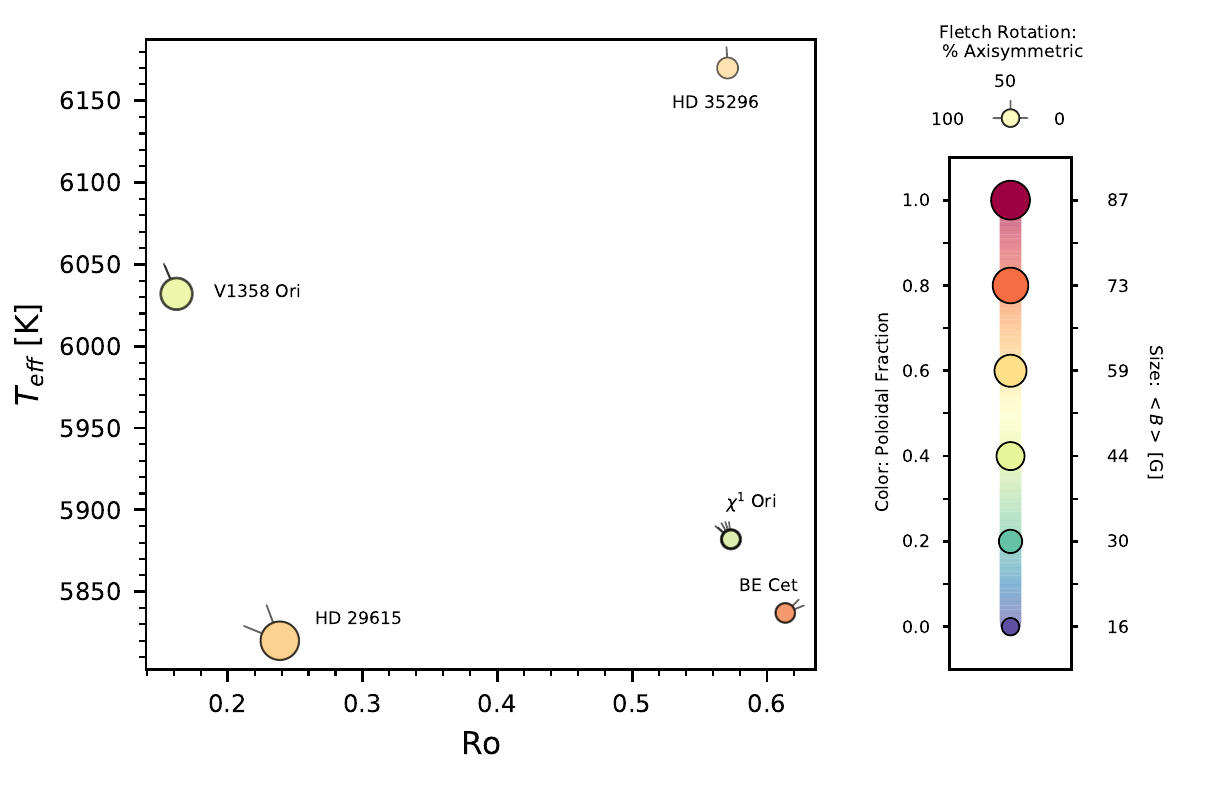}
   \caption{Effective temperature of the stars as a function of the Rossby number. The colour of the symbol corresponds to the fraction of poloidal field, and the size to the average surface magnetic field strength. The rotated fletches show the percentage of axisymmetric field in the map. For individual stars, variations of axisymmetry between different epochs are visually most clearly presented, but variations in the magnetic field strength can also in principle be seen as differences in the sizes of overplotted circles and variations in the poloidal fraction in their blended colours. In addition to our results from Table \ref{feat}, there is one additional BE Cet data point  and four $\chi^1$ Ori data points  from \cite{rosen16}, and one HD 29615  data point from \cite{hackman16}. See Sect. \ref{results} for more details.}
   \label{deconfusogram}
\end{figure}

\begin{figure}
   \centering
   \includegraphics[width=8cm]{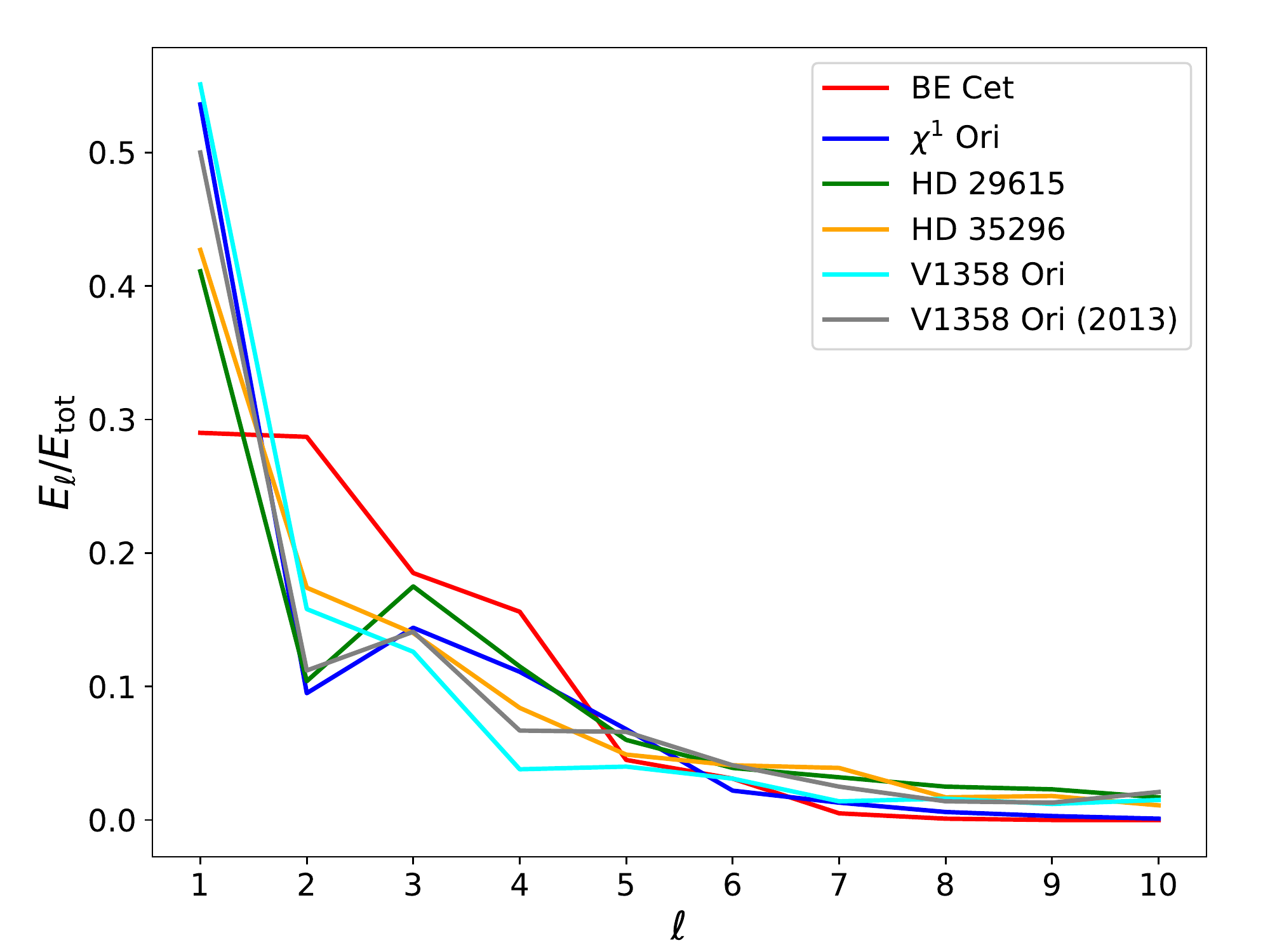}
   \caption{Distribution of the magnetic energy over the spherical harmonics.}
   \label{el}
\end{figure}

\subsection{Axisymmetry}

In Fig. \ref{deconfusogram} there is a potential trend of stars with lower Ro having more axisymmetric magnetic fields. V1358 Ori and HD 29615, the stars with the lowest Ro in our sample, have dominating axisymmetric components, while the rest of the stars have axisymmetric and non-axisymmetric components of roughly equal strength. In some of the earlier ZDI maps of \cite{rosen16}, however, $\chi^1$ Ori also has a fairly axisymmetric field.

\cite{viviani2018} found from numerical simulations that there is a transition from slowly rotating stars, where the magnetic field is dominated by the axisymmetric component, to more rapidly rotating stars, with a dominantly non-axisymmetric magnetic field, around the Coriolis number $\mathrm{Co} \approx 3$. With their definitions, Co is the inverse of Ro, so this would translate to $\mathrm{Ro} \approx 1/3 \approx 0.33$. The definition of $\tau_{\mathrm{conv}}$, however, is different in the context of simulations, where it is defined based on the volume averaged rms velocity and typical scales of turbulent eddies in the simulations, in contrast to the empirical relation from \cite{noyes84}, which is derived based on the stellar B-V colour. This transition region would be placed in between our axisymmetric, more active stars (HD 29615 and V1358 Ori, with $\mathrm{Ro} < 0.3$) and non-axisymmetric, less active ones (BE Cet, $\chi^1$ Ori and HD 35296, with $\mathrm{Ro} > 0.5$). The relation between Ro and the axisymmetry, however, is the opposite in our results compared to in theirs.

\cite{lehtinen16} found a similar transition as \cite{viviani2018} from photometrically observed stars, which they determined to be located in $\log R'_{\mathrm{HK}}$ space at $\log R'_{\mathrm{HK}} \approx -4.46$. 
In a later study, \cite{lehtinen2021} found two separate slopes in a $\log\mathrm{Ro}-\log R'_{\mathrm{HK}}$ diagram for main-sequence stars, where the knee point of the two slopes was located at $\mathrm{Ro} \approx 0.91$ in the scale of \cite{noyes84}. This knee point was identified as a potential transition region from axisymmetric to non-axisymmetric dynamo modes operating in the stars as it was located close to the transition point found by \cite{lehtinen16}. This is also close to the result of \cite{see2016}, where the transition between axisymmetric and non-axisymmetric solutions is seen around $\mathrm{Ro} \approx 1$.

In our results, however, the location of the possible transition region (around $\mathrm{Ro} \approx 0.3-0.6$) does not agree with \cite{lehtinen16,lehtinen2021} or \cite{see2016}. Also, the dependence of axisymmetry and Ro is inverted compared to the other studies. This could perhaps be explained with the small number of stars included in our study, but it is still an interesting feature.


\section{Conclusions}

We have continued to monitor the magnetic activity of a number of young, solar-type stars. In V1358 Ori, and possibly HD 35296, there have been polarity reversals between our observations and earlier published ones. Also, $\chi^1$ Ori seems to be an interesting target for future studies, with a possible rapid magnetic cycle.

For the stars with brightness maps, there is mostly a lack of any clear correlation between the spot structures and magnetic fields, as has been noted in earlier studies as well \citep[e.g.][]{hackman16,lehtinen2019arXiv}. Since most of the magnetic field goes undetected by ZDI \citep{kochukhov2020}, this could be explained if a large fraction of the magnetic field is actually composed of smaller structures with mixed polarities, which cancel one another out and thus go undetected in our maps.

Our sample includes three stars with a dominantly poloidal magnetic field (BE Cet, HD 29615, and HD 35296). In $\chi^1$ Ori, both components are close to equal, and V1358 Ori has a dominantly toroidal field. When comparing to previous studies of the same stars, there is variation, but as a general rule, there seems to be some tendency for the magnetic field of a certain star to stay poloidal or toroidal.

The magnetic energy distribution, seen in Fig. \ref{el}, shows that most of the magnetic energy is found in spherical harmonics of low order ($\ell \lesssim 5$). BE Cet stands out as the star with most energy in the orders $\ell \geq 2$, while the other stars have very similar distributions.

The deconfusogram in Fig. \ref{deconfusogram} shows a potential preference for a more axisymmetric 
magnetic field for stars with smaller Ro. For the poloidal versus toroidal field, this is not as clear, but there is a slight preference of a more toroidal field for the lower Ro. However, our sample is too small to draw any strong conclusions from this. A study with a larger sample of stars would have to be conducted to investigate this issue properly.

We also found that, in our cases, neglecting differential rotation did not have a significant effect on the results.

\begin{acknowledgements}

TW acknowledges the financial support from the Väisälä Foundation.
TH acknowledges the financial support from the Academy of Finland for the project SOLSTICE (decision No. 324161).
MJK acknowledges the support of the Academy of Finland ReSoLVE Centre of Excellence (grant No.~307411). This project has received funding from the European Research Council under the European Union's Horizon 2020 research and innovation programme (project "UniSDyn", grant agreement n:o 818665). 
OK acknowledges support by the Swedish Research Council, the Royal Swedish Academy of Sciences and the Swedish National Space Agency. 
SVJ acknowledges the support of the DFG priority program SPP 1992 "Exploring the Diversity of Extrasolar Planets" (JE 701/5-1).
\end{acknowledgements}

%
   \bibliographystyle{aa} 
   \bibliography{ESO} 
%



\begin{appendix}

\section{LSD profiles} \label{append1}

\begin{figure*}
   \centering
   \includegraphics[bb=30 50 400 600,width=6cm]{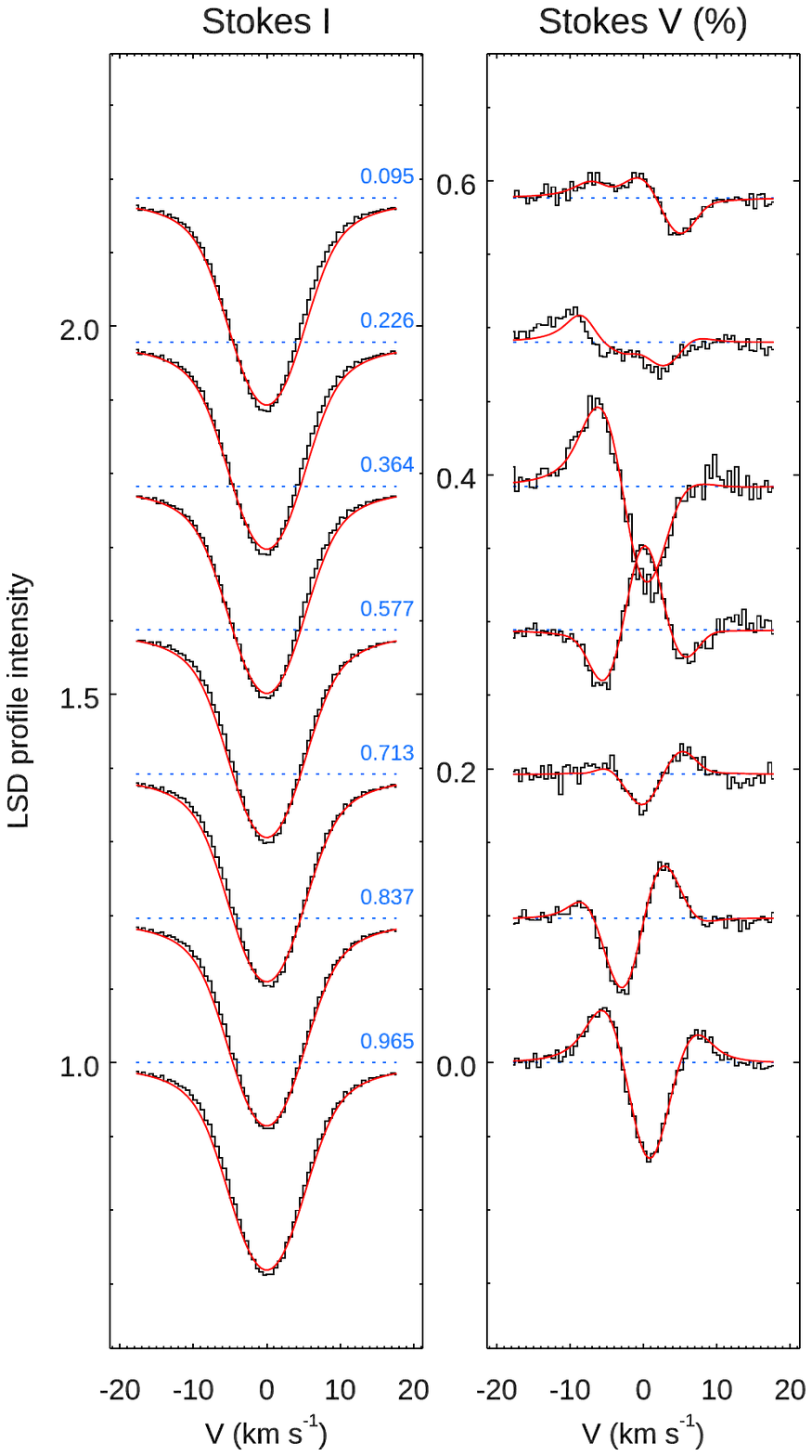}
   \includegraphics[bb=30 50 400 600,width=6cm]{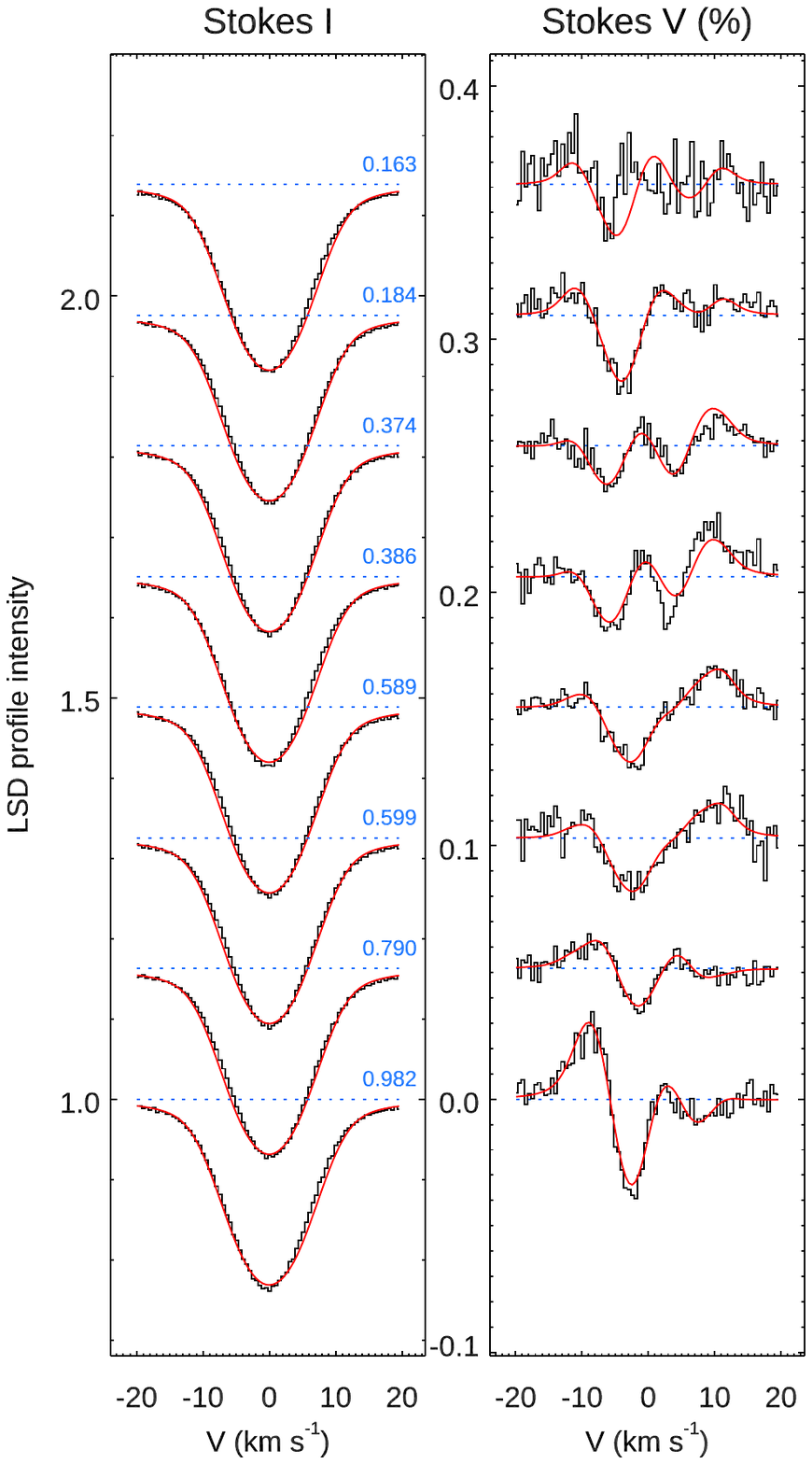}
   \includegraphics[bb=30 50 400 600,width=6cm]{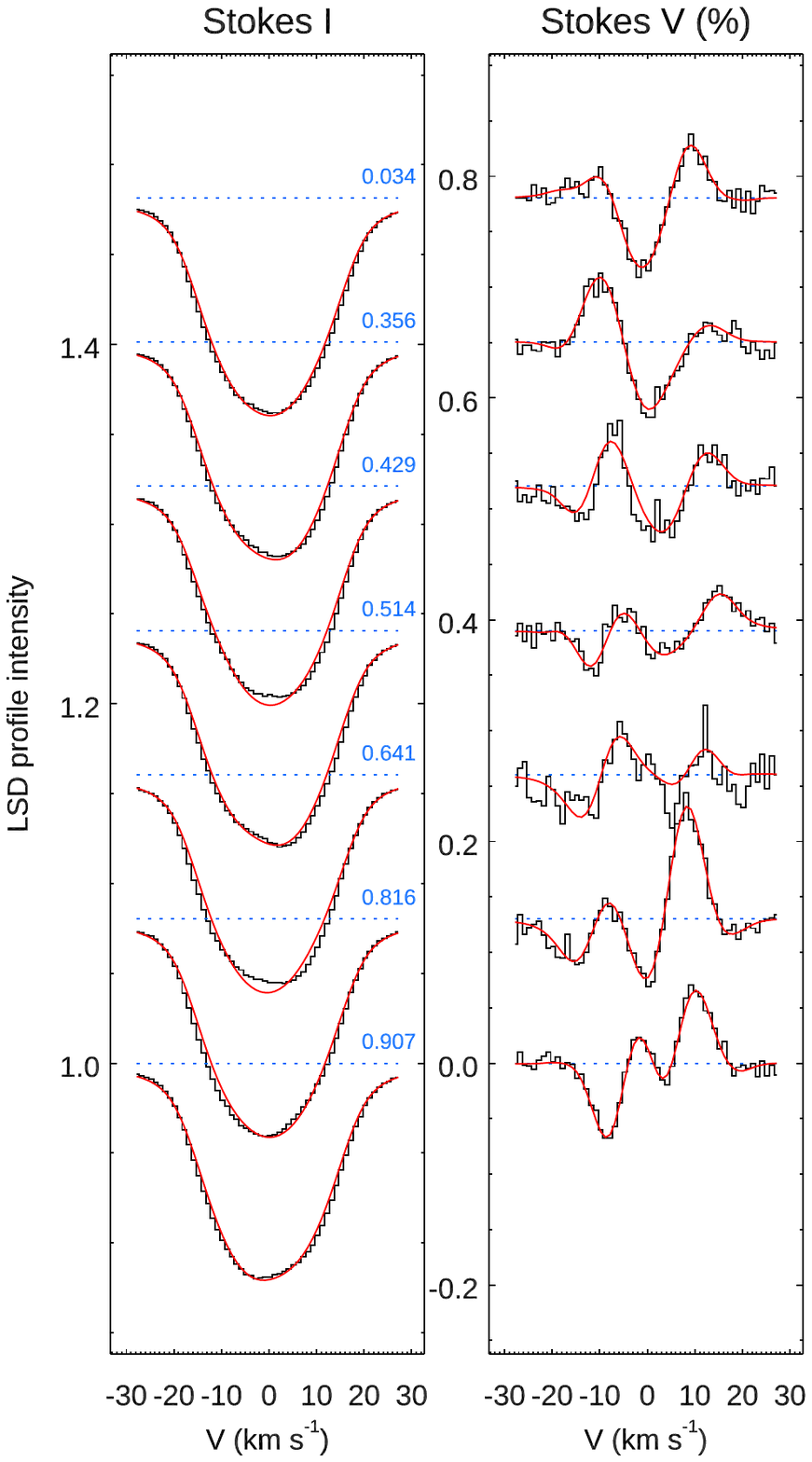}
   \includegraphics[bb=30 50 400 600,width=6cm]{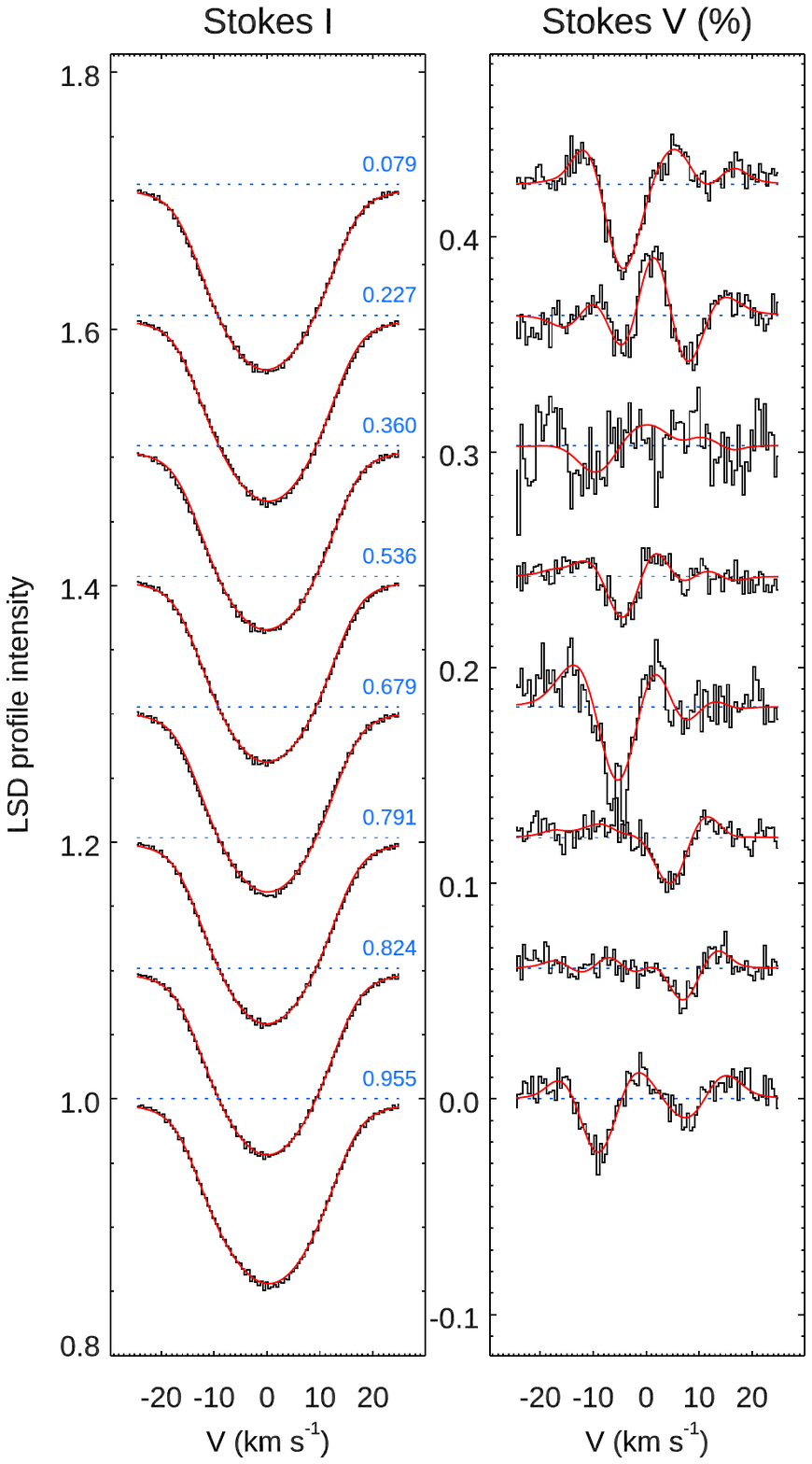}
   \includegraphics[bb=30 50 400 600,width=6cm]{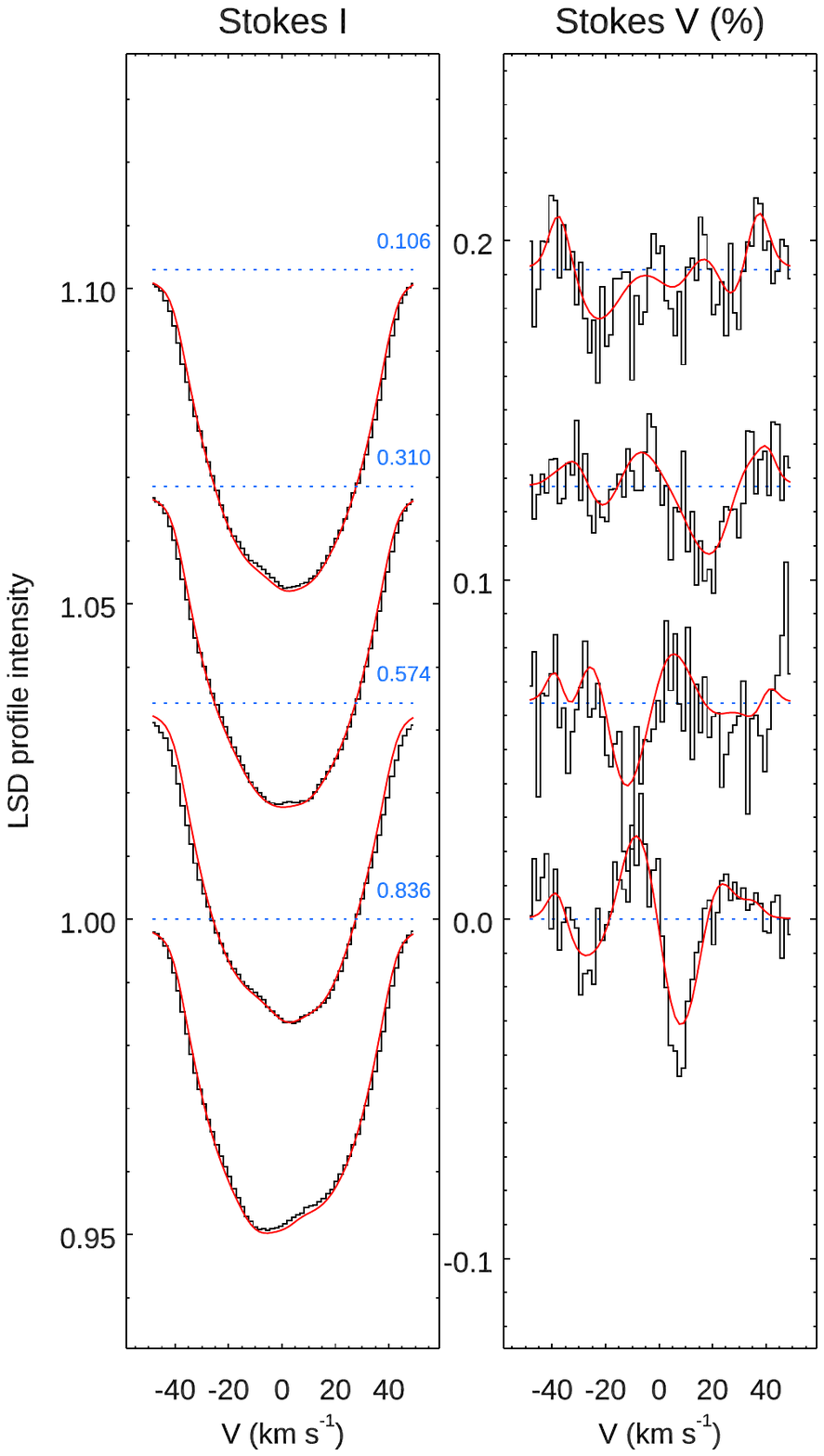}
   \includegraphics[bb=30 50 400 600,width=6cm]{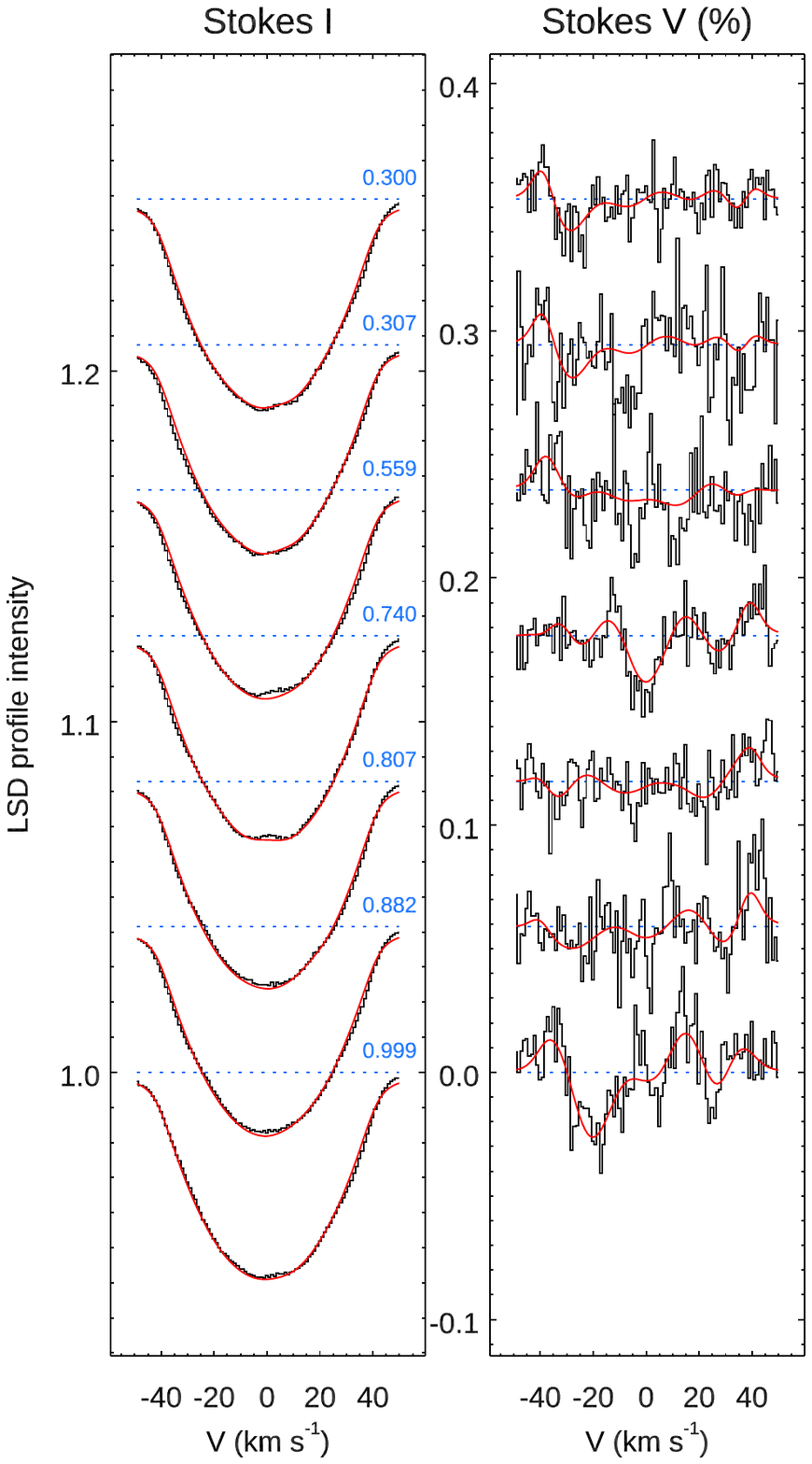}
   \caption{LSD profiles for BE Cet (upper left), $\chi^1$ Ori (upper centre), HD 29615 (upper right), HD 35296 (lower left), and V1358 Ori (lower centre - 2013; and lower right - 2017). The black line profiles are the observed Stokes profiles, and the red fits are the LSD profiles. The rotational phase is indicated for each profile.}
   \label{profs}
\end{figure*}

\section{LSD null profiles} \label{append2}

\begin{figure*}
   \centering
   \includegraphics[width=8cm]{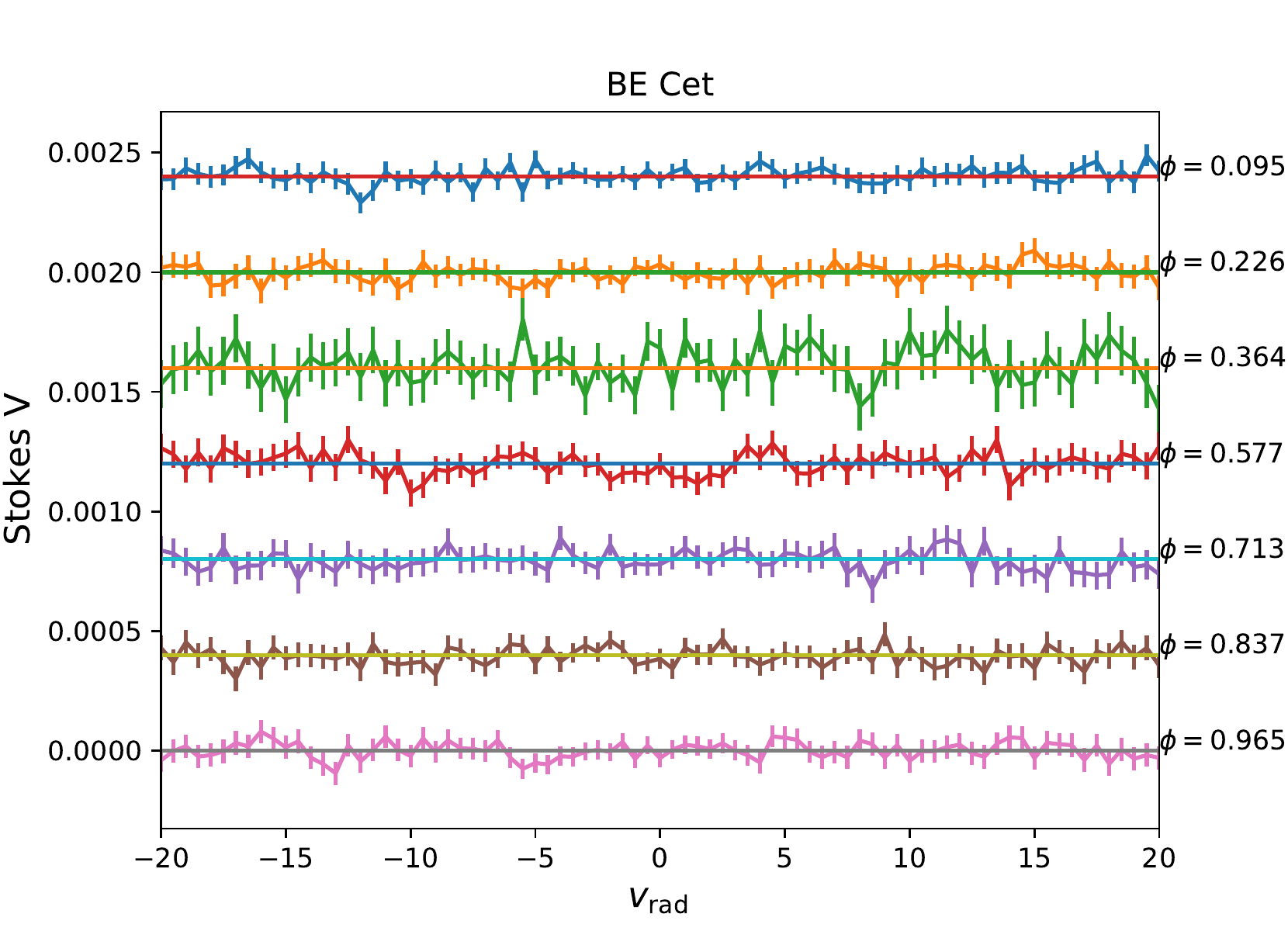}
   \includegraphics[width=8cm]{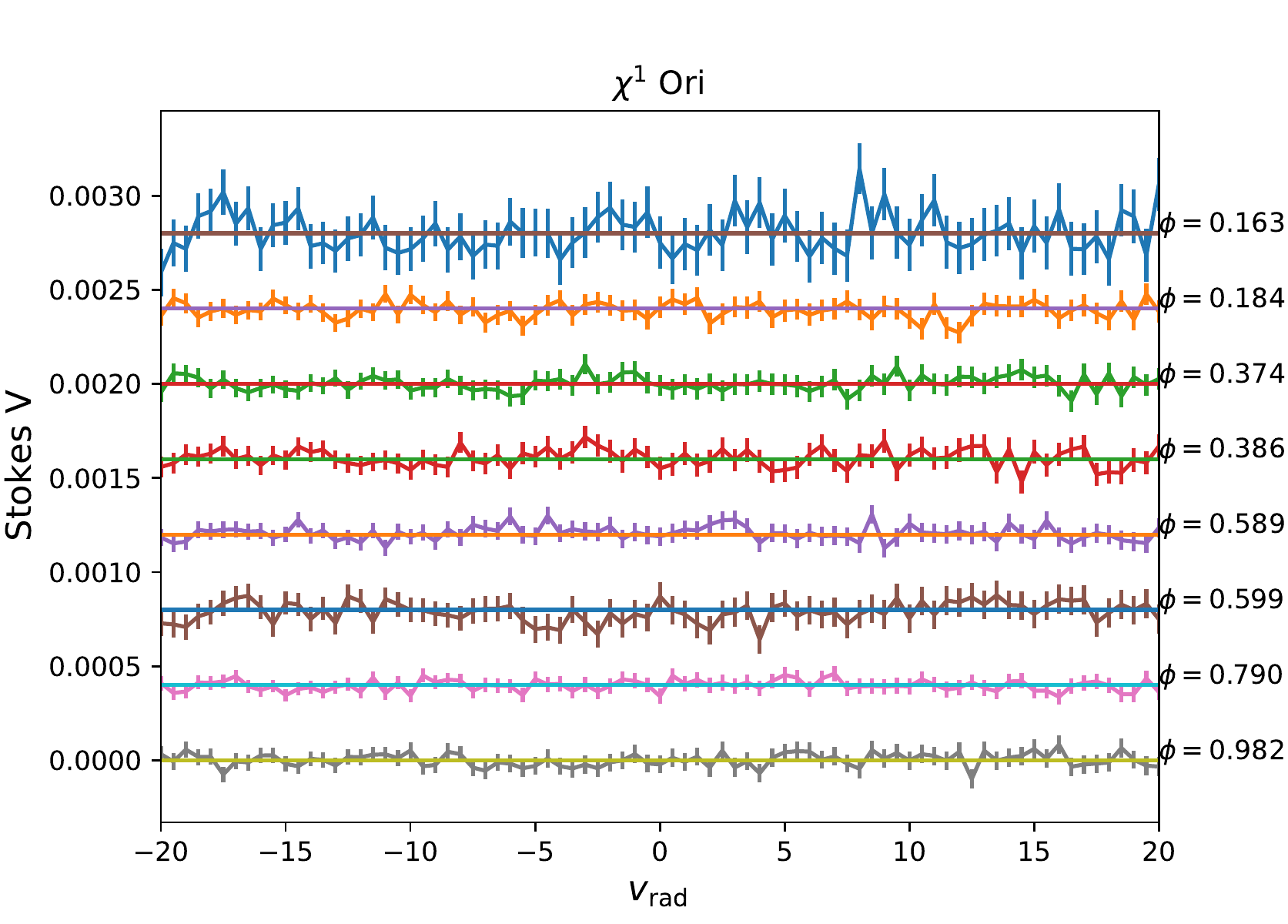}
   \includegraphics[width=8cm]{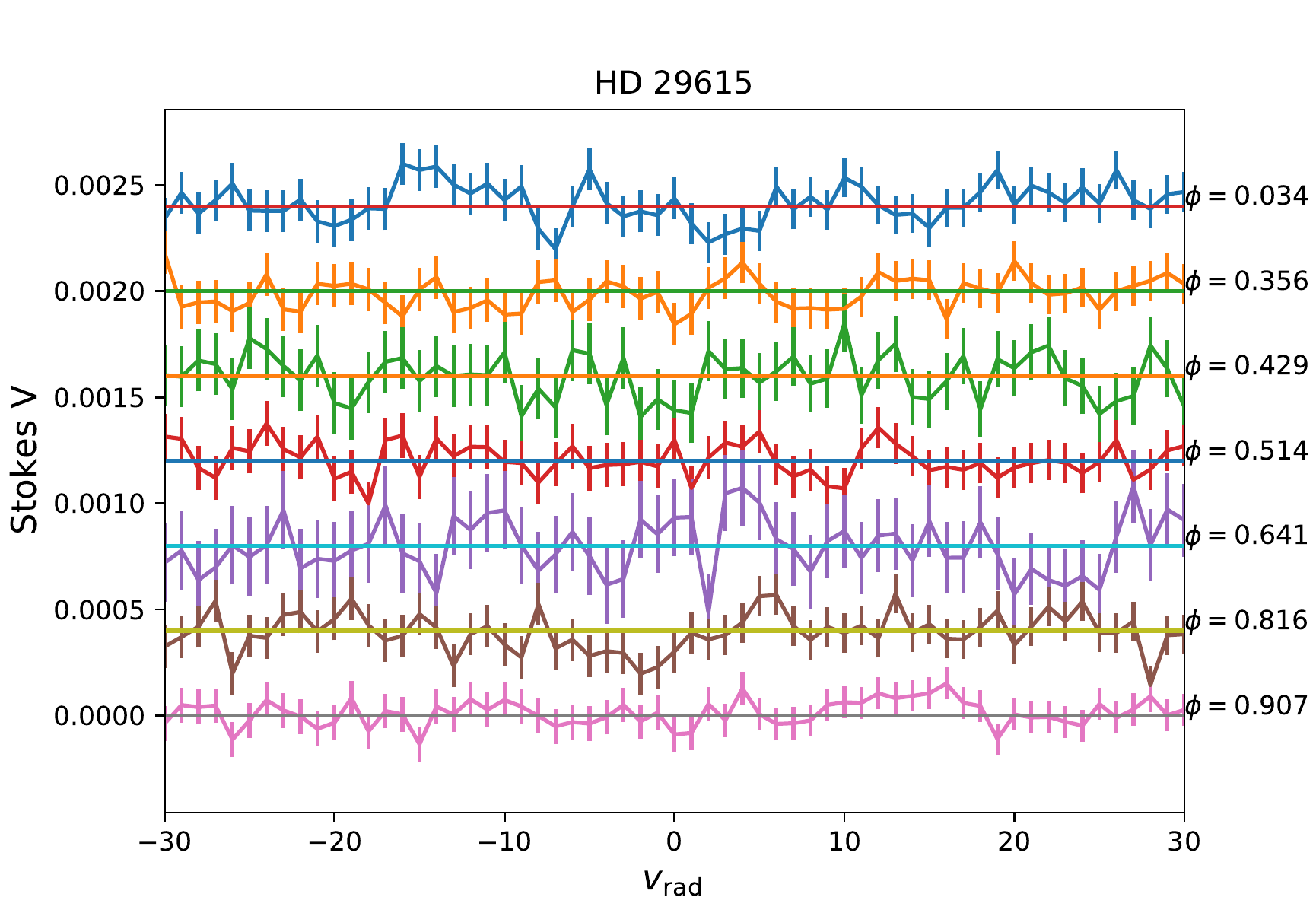}
   \includegraphics[width=8cm]{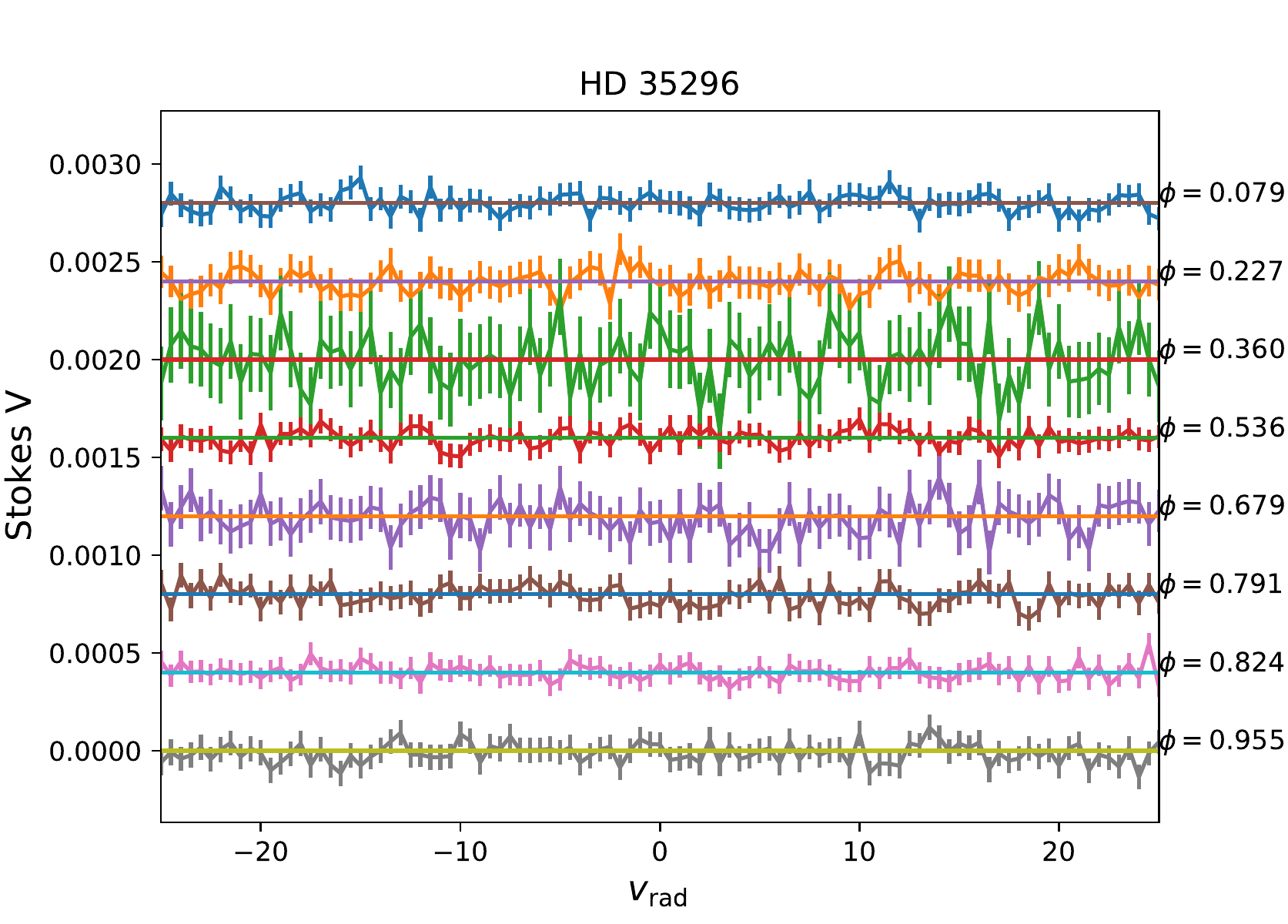}
   \includegraphics[width=8cm]{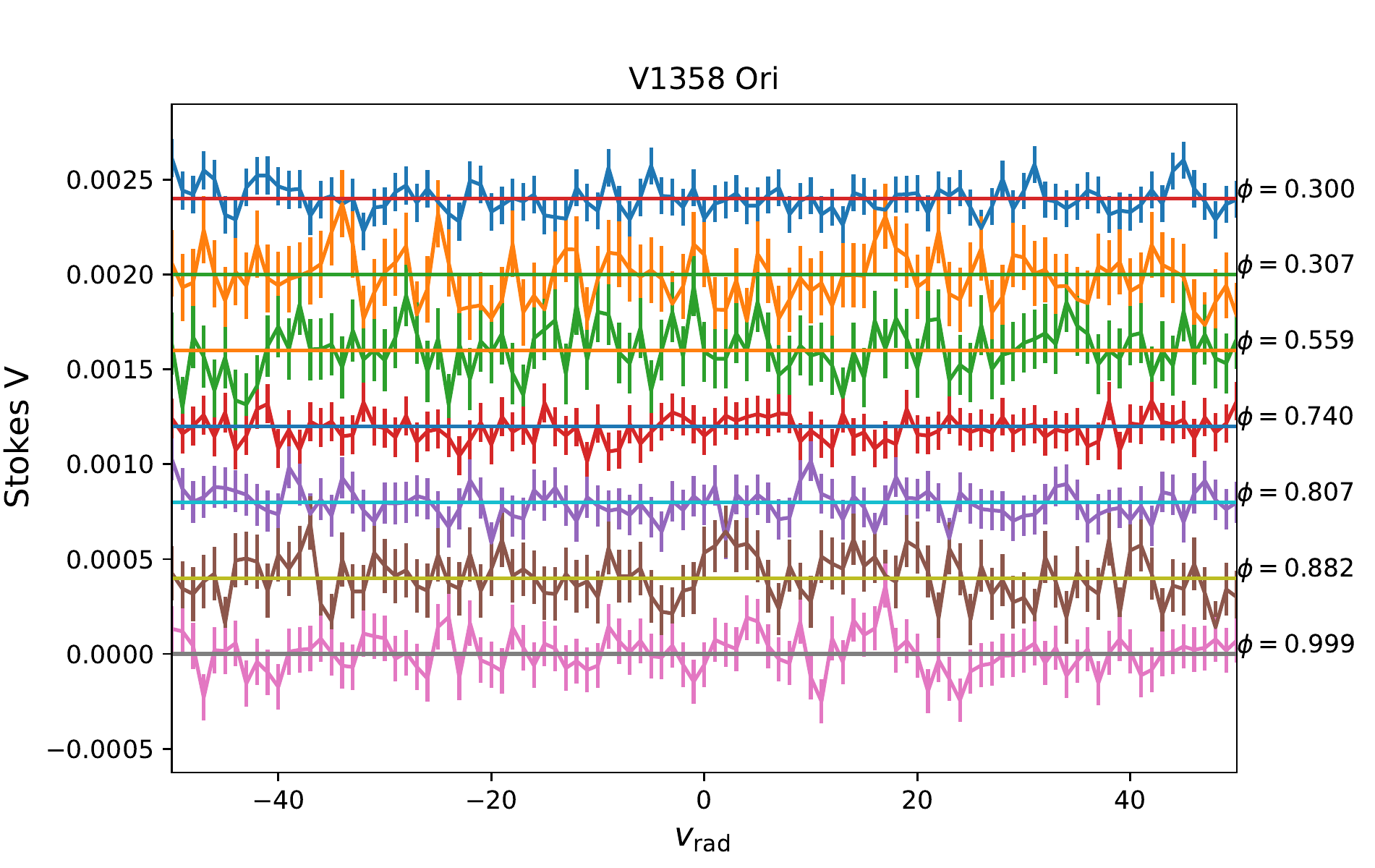}
   \caption{Stokes V null profiles for our stars. The rotational phase is indicated for each spectrum. The vertically shifted zero levels are shown as flat lines.}
   \label{null}
\end{figure*}

\section{Alternative ZDI maps with non-zero differential rotation: HD 29615} \label{append}

\begin{figure*}
   \centering
   \includegraphics[bb=30 200 800 340,width=\textwidth,angle=180]{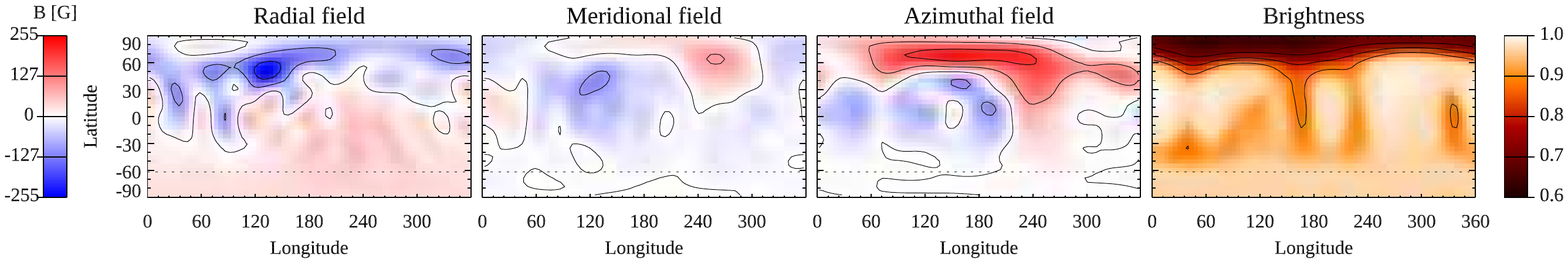}
   \includegraphics[bb=30 50 400 600,width=6cm]{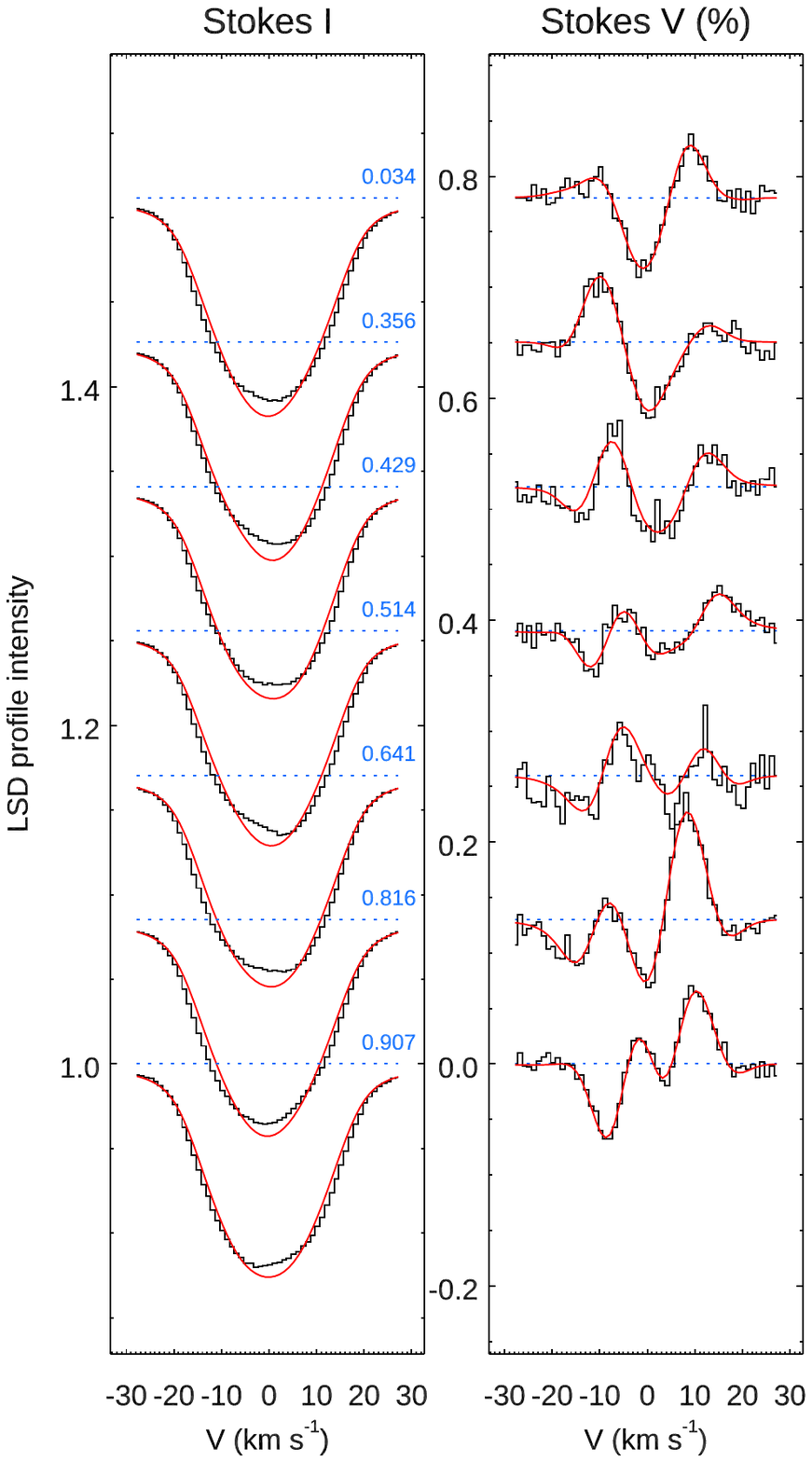}
   \caption{Alternative ZDI maps and LSD profiles for HD29615, using the differential rotation parameter $\alpha = 0.177$ and rotation period $P_{\mathrm{rot}} = 2.23$ d.}
   \label{hd29615_difrot}
\end{figure*}

\section{Alternative ZDI maps with non-zero differential rotation: V1358 Ori} \label{append_v1358ori}

\begin{figure*}
   \centering
   \includegraphics[bb=30 200 800 340,width=\textwidth,angle=180]{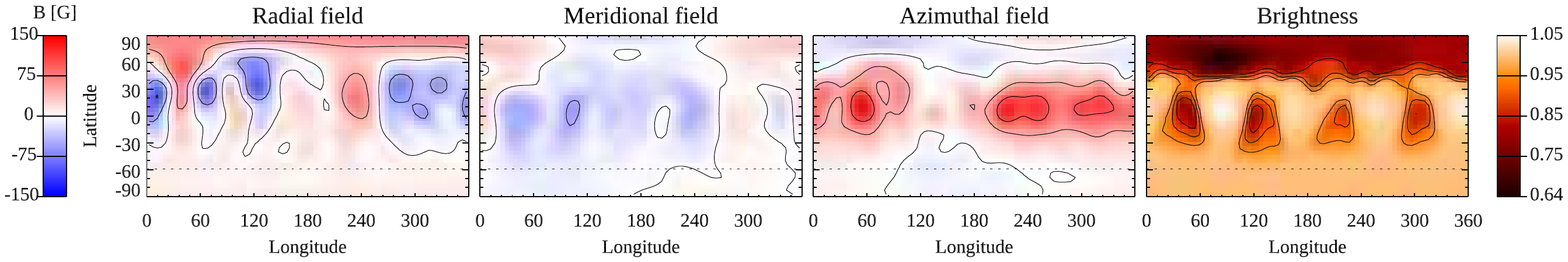}
   \includegraphics[bb=30 200 800 340,width=\textwidth,angle=180]{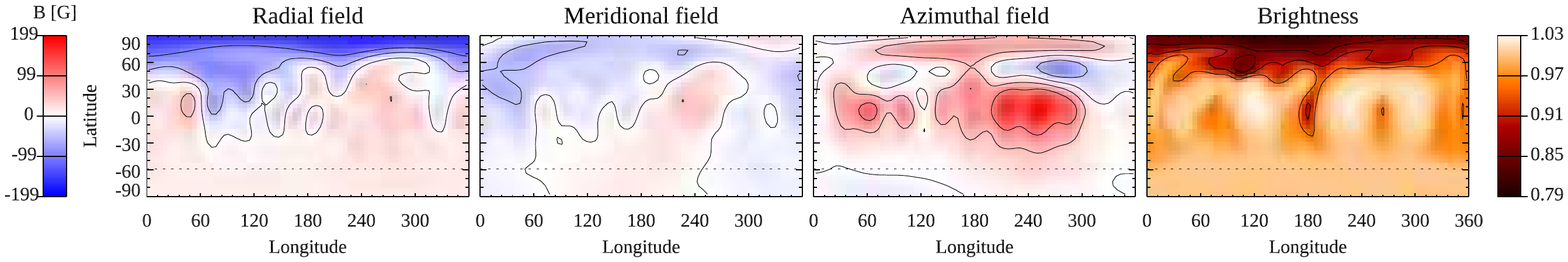}
   \includegraphics[bb=30 50 400 600,width=6cm]{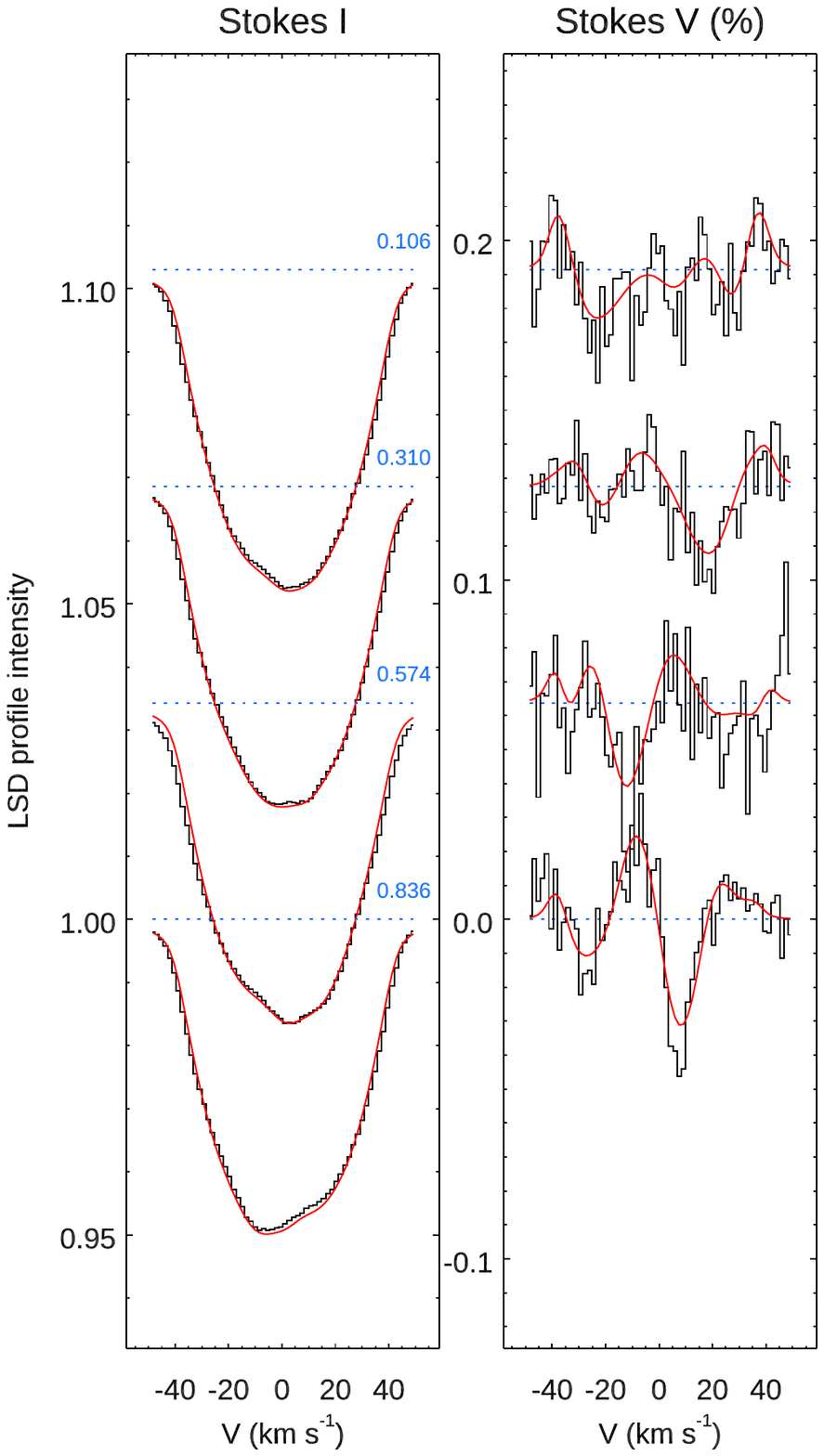}
   \includegraphics[bb=30 50 400 600,width=6cm]{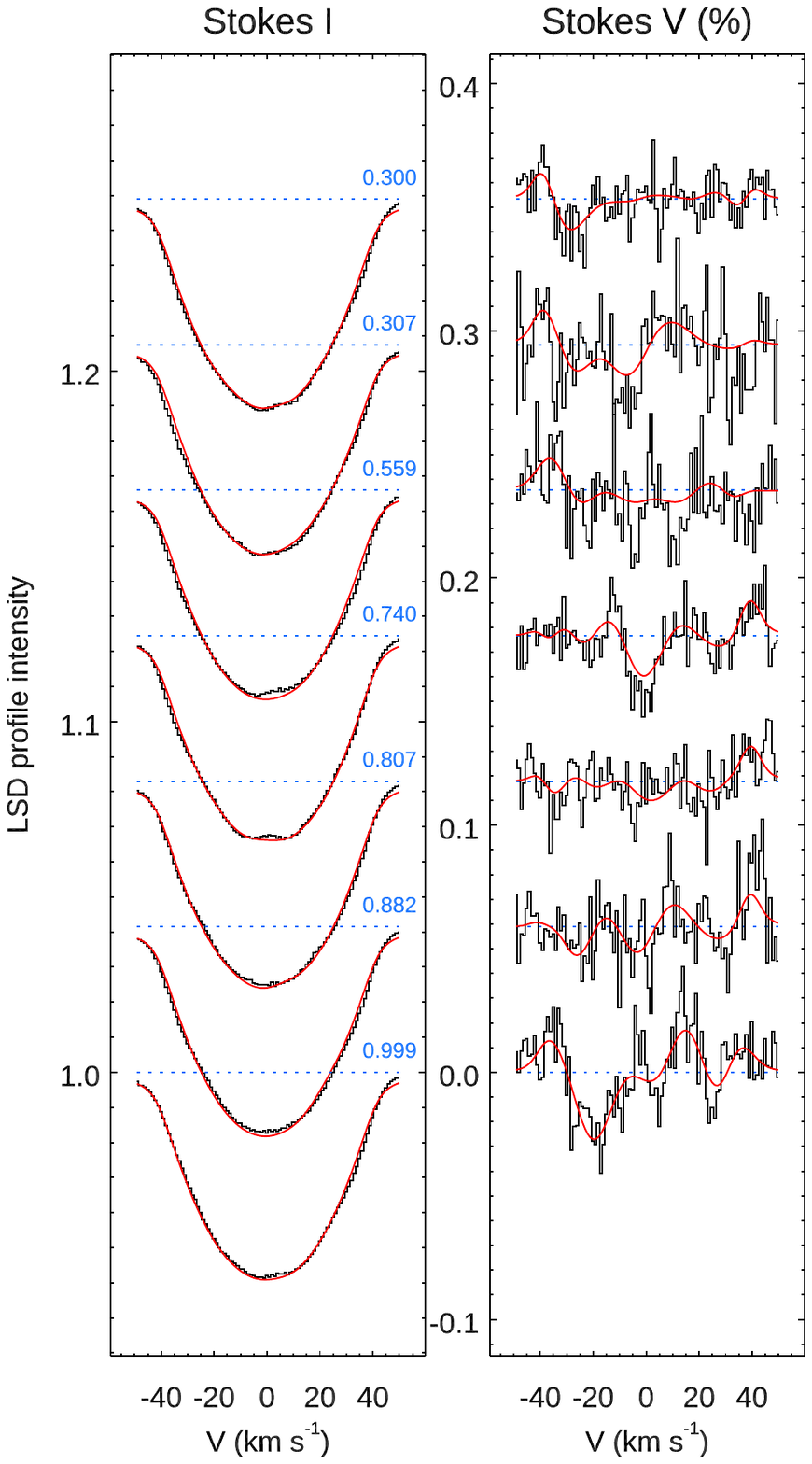}
   \caption{Alternative ZDI maps and LSD profiles for V1358 Ori, for 2013 (upper map, left line profiles) and 2017 (lower map, right line profiles), using the differential rotation parameter $\alpha = 0.016$ and rotation period $P_{\mathrm{rot}} = 1.335$ d.}
   \label{v1358ori_difrot}
\end{figure*}

\section{Additional parameters of the ZDI maps} \label{append4}

\begin{table*}
\centering
\caption{Additional parameters of the magnetic fields in the ZDI maps.}
\label{feat2}
\begin{tabular}{c c c c c c}
\hline\hline
Star & $\langle |B| \rangle_{\mathrm{rad}}$ [G] & $\langle |B| \rangle_{\mathrm{mer}}$ [G] & $\langle |B| \rangle_{\mathrm{azi}}$ [G] & Axis. ($m=0$) [\%] & Non-axis. ($m \neq 0$) [\%] \\
\hline
BE Cet & 10 & 5 & 9 & 23 & 77 \\
$\chi^1$ Ori & 6 & 4 & 9 & 49 & 51 \\
HD 29615 & 43 & 29 & 45 & 51 & 49 \\
HD 35296 & 10 & 9 & 13 & 44 & 56 \\
V1358 Ori (2013) & 23 & 15 & 35 & 57 & 43 \\
V1358 Ori (2017) & 23 & 15 & 44 & 59 & 41 \\
\hline
\end{tabular}
\end{table*}

\section{Fraction of magnetic energy in spherical harmonic expansions} \label{append3}

\begin{table*}
\centering
\caption{Fraction of magnetic energy (in \%) in each spherical harmonic expansion.}
\label{el_frac}
\begin{tabular}{c c c c c c c c c c c}
\hline\hline
Star & $\ell=1$ & $\ell=2$ & $\ell=3$ & $\ell=4$ & $\ell=5$ & $\ell=6$ & $\ell=7$ & $\ell=8$ & $\ell=9$ & $\ell=10$ \\
\hline
BE Cet & 29.0 & 28.7 & 18.5 & 15.6 & 4.5 & 3.1 & 0.5 & 0.1 & 0.0 & 0.0 \\
$\chi^1$ Ori & 53.6 & 9.5 & 14.4 & 11.1 & 6.8 & 2.2 & 1.3 & 0.6 & 0.3 & 0.1 \\
HD 29615 & 41.1 & 10.4 & 17.5 & 11.5 & 6.0 & 3.9 & 3.2 & 2.5 & 2.3 & 1.7 \\
HD 35296 & 42.7 & 17.4 & 14.0 & 8.4 & 4.9 & 4.1 & 3.9 & 1.7 & 1.8 & 1.1 \\
V1358 Ori (2013) & 50.0 & 11.2 & 14.1 & 6.7 & 6.6 & 4.1 & 2.5 & 1.4 & 1.3 & 2.1 \\
V1358 Ori (2017) & 55.1 & 15.8 & 12.6 & 3.8 & 4.0 & 3.1 & 1.4 & 1.6 & 1.2 & 1.5 \\
\hline
\end{tabular}
\end{table*}

\end{appendix}

\end{document}